\documentclass[%
 reprint,
 amsmath,amssymb,
 aps,
]{revtex4-2}

\usepackage{graphicx}
\usepackage{dcolumn}
\usepackage{bm}
\usepackage{hyperref}

\begin{document}

\preprint{APS/123-QED}

\title{Detectivity and bandwidth limits of cooled and uncooled \\ light detection using nanomechanical resonators}

\author{Mathis Turgeon-Roy}
 \altaffiliation[Also at ]{Department of Computer Science and Software Engineering, Laval University, Québec G1V 0A6}
\author{Mohammed Shakir}%
\author{Zachary Louis-Seize}
\author{Raphael St-Gelais}
 \altaffiliation[Also at ]{Department of Physics, University of Ottawa, Ottawa, Ontario K1N 6N5, Canada}
 \email{raphael.stgelais@uottawa.ca}
\affiliation{%
Department of Mechanical Engineering, University of Ottawa, Ottawa, Ontario K1N 6N5, Canada\\
}%
\date{\today}

\begin{abstract}
Nanomechanical resonators (NMRs) offer a promising alternative to traditional thermal-based radiation detectors due to their immunity to electrical noise. In recent years, these sensors have reached the previously unattained theoretical detectivity limit set by the fluctuation noise of thermal photons at room temperature. Beyond this point, improvements of NMR resonators do not translate into greater detectivity, but in greater effective bandwidth. There is, however, no simple model predicting the limits of this bandwidth enhancement. Likewise, models predicting the performances of NMR-based radiation sensors under active cooling have not been derived. To address these gaps in knowledge, a key missing ingredient consists of defining the NMR optimal driven amplitude that minimizes additive frequency noise, but without performance degradation from nonlinear phenomena. We find that, in the context of NMR-based radiation sensing, this optimal amplitude ($a_\mathrm{opt}$) is dramatically different than the commonly assumed critical amplitude ($a_\mathrm{c}$) that defines the onset of non-linear phenomena in nanomechanical resonators. Our proposed model for this optimal amplitude allows us to quantify the maximum bandwidth enhancement in NMR-based radiation sensors. We also derive simple equations predicting the maximum detectivity in cryogenically cooled sensors. Finally, combining these two models allows us to define new universal performance limits. We find that the maximum weighted product of detectivity and bandwidth,  $D_T^{*\:3/2} \times \omega_\mathrm{BW}$, is remarkably simple and depends mostly on intrinsic resonator material properties and dimensions. This unveils important general conclusions on the ideal geometry of NMR radiation sensors. We find that thermomechanically-limited sensors should be as thin and extended as possible. In contrast, readout-limited sensors should also be thin, but should be just large enough to make radiation heat transfer dominant compared to conduction.


\end{abstract}

\maketitle


\section{\label{sec:level1}Introduction\protect\\ }
Radiation sensors such as resistive bolometers \cite{wang_vanadium_2012,renoux_subwavelength_2011}, pyroelectric detectors \cite{hossain_pyroelectric_1991,liu_pyroelectric_1978} and thermopiles \cite{hsu_graphene_2015,stantoine_nanotube_2011} rely on electrical temperature sensing and therefore suffer from electrical noise that prevent them from reaching the fundamental limit of thermal-based radiation sensing. In contrast, nanomechanical resonator-based  sensors \cite{zhang_nanomechanical_2013,laurent_electromechanical_2018,blaikie_graphene_2019,hui_plasmonic_2016,piller_thermal_2023,zhang_high_2024,zhang_room_2016,vicarelli_micromechanical_2022,He2026} detect temperature mechanically and can therefore be immune to electrical noise, for example by employing laser-based readouts. This has allowed performances reaching the fundamental detectivity ($D^*$) limit of thermal photon fluctuations  in recent work  \cite{zhang_enhanced_2025,kanellopulos_comparative_2025}.

 After reaching the temperature fluctuation noise limit, improvements in mechanical and readout noises do not translate into higher detectivity, but into increased sensing bandwidth ($\omega_{\mathrm{BW}}$) \cite{zhang_enhanced_2025}. However, current models predicting the maximum value of this bandwidth enhancement are incomplete. Frequency instability due to thermomechanical and transduction noise are well predicted by models accounting for closed-loop frequency tracking \cite{demir_understanding_2021, Schmid2023Fundamentals, zhang_enhanced_2025}, but those are limited to the linear oscillation regime and do not predict the maximum possible drive amplitude of the NMR. Likewise, recent work \cite{zhang_enhanced_2025} demonstrated that operating well beyond the commonly accepted nonlinear critical drive amplitude \cite{Schmid2023Fundamentals,lifshitz_nonlinear_2008} can improve the frequency stability of NMR-based radiation sensors. The optimal drive amplitude value in the context of NMR-based radiation sensing therefore remains unknown, thus preventing accurate predictions of the maximum sensing bandwidth. 
 
To predict the optimal drive amplitude ($a_\mathrm{opt}$) in NMR-based radiation sensor, we derive an expression for frequency noise resulting from thermomechanical fluctuations in a non-linear actuation regime. The model accounts for both Duffing nonlinearity in closed loop frequency tracking \cite{manzaneque_resolution_2023}, and for amplitude-dependent damping \cite{catalini_modeling_2021}. We then put this consolidated noise model in perspective with the fundamental temperature fluctuations noise \cite{zhang_enhanced_2025} that affect NMR-based radiation sensors. 
 
Similarly, it is a well known fact that cooling thermal-based detectors can improve sensing performances \cite{mather_bolometer_1982,kruse_elements_1962}, but models for performance under active cooling have not been derived in the context of nanomechanical radiation sensors. We therefore adapt heat transfer models describing power fluctuations ($S_P$) from multibody radiative exchange  \cite{kruse_elements_1962} to the particular case of NMRs radiation sensors in cooled enclosures.  

Building on these models for detectivity limits at various temperatures (Sec. II) and nonlinear bandwidth limits (Sec. III), we propose a universal figure of merit for predicting the maximum performance of damping-diluted NMR radiation sensors. More specifically, we find that the maximum product of $\left(D^{*}\right)^{3/2}$ and $\omega_{\mathrm{BW}}$ in NMR-based sensor is remarkably simple, depending essentially on intrinsic material parameters, while being independent of power fluctuations in the sensor, making it a simple and widely applicable performance ceiling.

\section{Detectivity Limits}

The detectivity ($D^*$) (in $\mathrm{cm\sqrt{Hz}/W}$) of a sensor quantifies its ability to detect small signals and distinguish them from its internal noise. It is defined as
\begin{equation}\label{Detectivity1}
D^{*} = \frac{A^{1/2}}{\mathrm{NEP}} ,
\end{equation}
where ${A}$ is the sensing area. When performing thermal-based radiation sensing with frequency shift-based sensors, noise equivalent power (NEP) (in $\mathrm{W} \,\mathrm{Hz}^{-1/2}$) is given by
\begin{equation}\label{eqNEP}
\mathrm{NEP} = \frac{S_{\mathrm{y}}^{1/2}}{R},
\end{equation}
where ${S_\mathrm{y}}$ quantifies the normalized ($y=\Delta f/f$) frequency ($f$) fluctuation noise density (single-sided) of the resonator expressed in $\mathrm{Hz}^{-1}$, while ${R}$ (in $\mathrm{W}^{-1}$) represents the responsivity of normalized frequencies (i.e., of $y$) to radiation exposure and is given by
\begin{equation} \label{eqR}
R(\omega) = \frac{\gamma_{\mathrm{abs}}\,\alpha}{G}\,
\left| H_{\mathrm{th}}(\omega) \right|.
\end{equation}
In \autoref{eqR}, $\gamma_{\mathrm{abs}}$ is the optical absorption coefficient (unitless) evaluated at the detection wavelength, $\alpha$ is the thermal coefficient of fractional frequency shift (in $\mathrm{K}^{-1}$) and $G$ (in $\mathrm{W/K}$) denotes the differential thermal conductance linking the nanomechanical resonator (NMR) at temperature $T_d$ to its surrounding. This thermal conductance $G$ includes radiative coupling (through $G_{\mathrm{rad}}$) by two surfaces of area $A$, as well as solid state conduction ($G_{\mathrm{cond}}$):
\begin{equation}
G =  8\varepsilon_{T_d}\sigma_{SB} AT_d^3+G_{\mathrm{cond}}. \label{eq Gtot}
\end{equation}
In \autoref{eq Gtot}, $\varepsilon_{T_d}$ denotes the total hemispherical emissivity of the sensor surface evaluated at temperature $T_d$, and $\sigma_{SB}$ is the Stefan-Boltzmann constant. $G_{\mathrm{cond}}$ depends heavily on the sensor geometry. Expressions are given in \cite{kanellopulos_comparative_2025} for drumheads, strings and trampolines. In the simplest possible case of a plain 2D membrane resonator, this term reduces approximately to $G_{\mathrm{cond}}=8\pi h k$, \cite{kanellopulos_comparative_2025} where $h$ and $k$ are respectively the membrane thickness and its material thermal conductivity. Note that many important limit performance cases occur when $G_{\mathrm{cond}}\approx0$. We therefore often set $G\approx G_{\mathrm{rad}}$ when this allows for significant limit case simplifications.

In turn, the filter $H_{\mathrm{th}}(\omega)$ captures the thermal response of the device as
\begin{equation}
H_{\mathrm{th}}(\omega) = \frac{1}{1 + j\omega\tau_{\mathrm{th}}},
\end{equation}
where, $\tau_{\mathrm{th}}=C/G$ is the sensor characteristic thermal response time for a given heat capacity $C$ (in J/K).

The fractional frequency noise (${S_\mathrm{y}}$) in \autoref{eqNEP} encompasses both the fundamental temperature fluctuations  ($S_\mathrm{y,T}$) as well as additive noises that should be minimized as much as possible to increase sensor performances. Readout noise ($S_\mathrm{y{,read}}$) and  thermomechanical noise ($S_\mathrm{y\mathrm{,thm}}$) are typically the most significant of these additive noise sources \cite{Schmid2023Fundamentals}, such that we express the total frequency noise as:
\begin{equation}
S_\mathrm{y} = S_\mathrm{y,T} + S_\mathrm{y\mathrm{,thm}} + S_\mathrm{y\mathrm{,read}}.
\end{equation}
In an ideal scenario, all additive noise sources are minimized, such that $S_\mathrm{y}\approx S_\mathrm{y,T}$ and frequency noise is given by
\begin{equation}\label{Syth}
S_\mathrm{y,T} = 
\frac{S_P}{G^{2}}\,\alpha^{2}\,
\left| H_{\mathrm{th}}(\omega) \right|^{2} ,
\end{equation}
where $S_P$ quantifies the power fluctuations (in $\mathrm{W^2/Hz}$) resulting from random energy exchanges between the NMR and its thermal environment. Note that throughout this work, $S_\mathrm{y}$ expressions are always given for closed-loop frequency tracking with an infinite tracking bandwidth, as discussed in greater detail in \autoref{Section bandwidth}. 

\begin{figure}[t]
    \centering
    \includegraphics[width=1.0\linewidth]{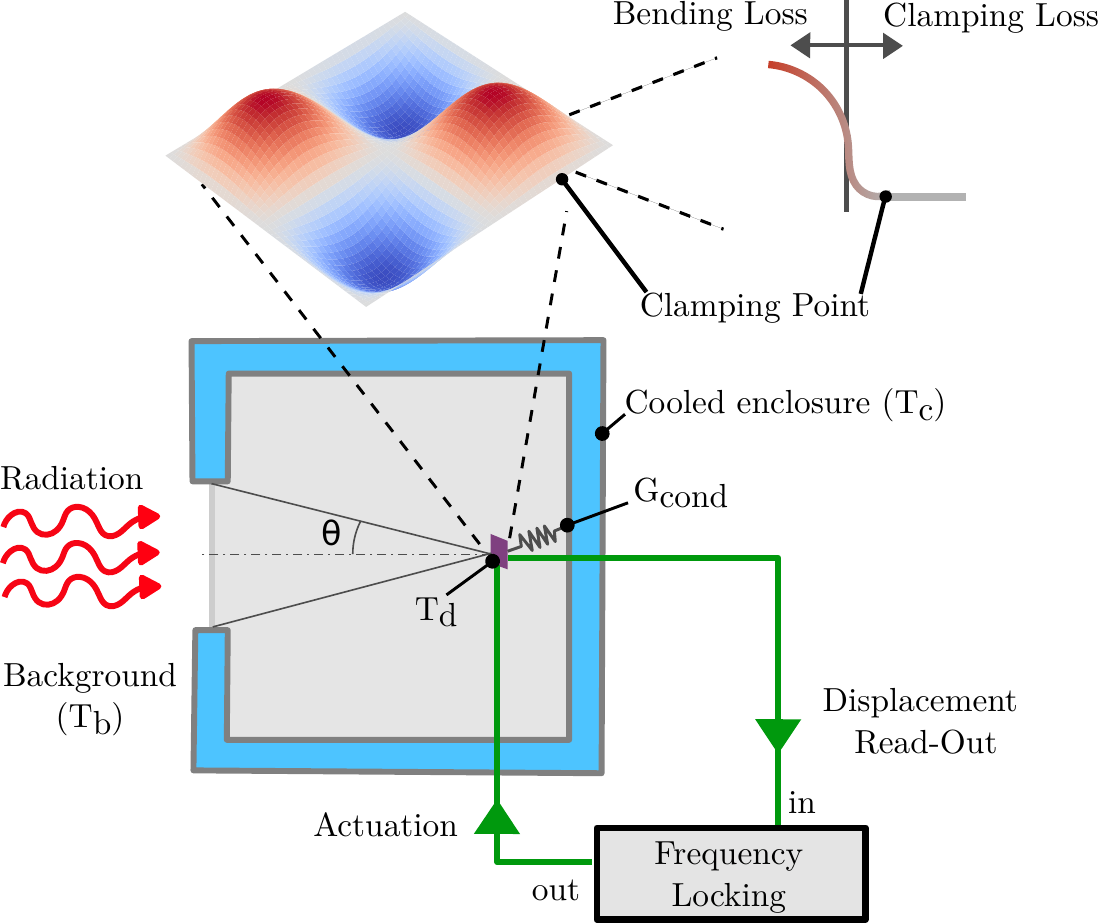}
    \caption{Schematic of the nanomechanical radiation sensing configuration considered in this work, where incoming radiation changes the eigenfrequency of a vibrating nanomechanical resonator (for example a square membrane vibrating at eigenmode order (2,2) in this image). The resonator temperature, $T_d$,  can be lower than the outside background temperature, $T_b$, when the casing is actively cooled to a temperature $T_c$. The detector and background are coupled radiatively through an opening of half angle $\theta$. Coupling between the detector and the casing occurs both radiatively, and through solid state heat conduction (through $G_{\mathrm{cond}}$).  In the square membrane example, damping of the mechanical vibration is accounted for by considering both mode bending loss and sharp bending at the clamping point (see top-right schematic).}
    \label{fig:Visualisation}
\end{figure}

\begin{figure}[!ht]
    \centering
    \includegraphics[width=\linewidth]{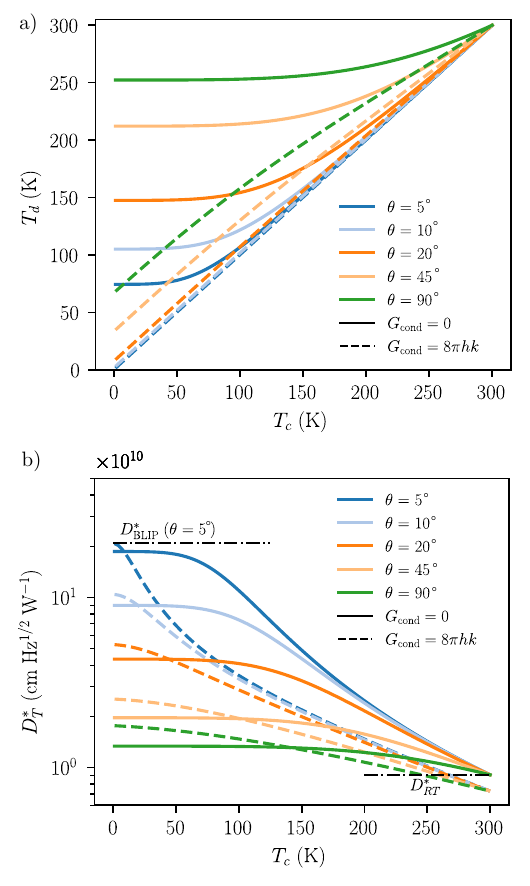}

    \caption{
    Detectivity limits of cooled detectors as a function of casing temperature $T_c$
    and opening angle $\theta$, considering a resonator radiatively coupled on both surfaces, the grey-body assumption
    $\varepsilon=\gamma_{\mathrm{abs}}=0.5$, and background temperature $T_b=300$~K.
    For $G_{\mathrm{cond}}=0$ (solid lines), the limits are independent of sensor area,
    geometry, or material. The effect of non-negligible conduction is illustrated
    using $G_{\mathrm{cond}}$ value of a typical square silicon nitride resonator
    (sensing area $A=1.44~\mathrm{mm}^2$, thickness $h=50$~nm, thermal conductivity $k=2.7$~W\,m$^{-1}$\,K$^{-1}$).
    (a) Temperature of the detector $T_d$ as a function of casing temperature $T_c$
    for several opening half-angles $\theta$.
    (b) Detectivity $D_T^{*}$ as a function of casing temperature $T_c$ and opening
    half-angles $\theta$. At low temperatures, the detectivity approaches the
    background-limited value $D^{*}_{\mathrm{BLIP}}$.
    }
    \label{fig:Detectivity_vs_Tc}
\end{figure}

In a typical radiation sensor geometry (see \autoref{fig:Visualisation}) the NMR sensor is contained in a casing at temperature $T_c$, that comprises an opening (with a half angle $\theta$) through which the detector is exposed to a background at temperature $T_b$. We assume that the sensor is coupled to its surroundings (i.e., the background and casing) through both its front and back surfaces (e.g., a freestanding thin film).  In steady-state, the net energy exchange is null between the detector and two thermal baths (at  $T_b$ and $T_c$). This yields the energy balance equation given in Appendix \ref{Steady state eq}, from which we can obtain $T_d$ (see \autoref{fig:Detectivity_vs_Tc} (a)). In the limit case where $G_\mathrm{cond}\approx0$, this reduces to a simple analytical expression:
\begin{equation} \label{eq Td}
T_{d} =
\left(
\frac{
\varepsilon_{T_b}T_{b}^{4}\sin^{2}\theta
+ \varepsilon_{T_c}T_{c}^{4}\left(1+\cos^{2}\theta\right)
}{2 \varepsilon_{T_d}}
\right)^{1/4} .
\end{equation}

In \autoref{eq Td}, $\varepsilon_{T}$ denotes the total hemispherical emissivity of the detector surface evaluated at the different system temperatures. In most cases, it is acceptable to assume a weak temperature dependence (i.e., grey body assumption), such that $\varepsilon_{T_b}\approx\varepsilon_{T_c}\approx\varepsilon_{T_d}\equiv\varepsilon$. 

After calculating the detector temperature, the thermal power fluctuations due to energy exchanges between the sensor and the different thermal baths can be calculated from formalisms given in \cite{kruse_elements_1962} (for the radiative term in the left hand side in \autoref{eq SW}) and in \cite{mather_bolometer_1982} for the conduction term (right hand side in \autoref{eq SW}): 
\begin{equation} \label{eq SW}
\begin{aligned}
S_P ={}& A \sigma_{SB} k_B \Big( 
16\,\varepsilon_{T_d} T_d^5 
+\, 8\,\varepsilon_{T_b} T_b^5 \sin^2\theta\\
&\quad
+\, 8\,\varepsilon_{T_c} T_c^5 (1+\cos^2\theta) 
\Big)
+ 2 k_B (T^2_d + T^2_c) G_{\mathrm{cond}} ,
\end{aligned}
\end{equation}
where ${k_B}$ is the Boltzmann constant. Note that the last term (i.e. conduction) in \autoref{eq SW} is a common approximation in the field of cooled bolometer \cite{misiak_developpements_2021} that considers the sensor and the thermal bath $T_b$ as coupled through a lumped solid state conduction $G_{\mathrm{cond}}$. More accurate expressions can be obtained by integrating conduction over the sensor geometry as in \cite{mather_bolometer_1982}. The simple approximation of \autoref{eq SW} is nevertheless sufficient in the present case since our goal is generally to minimize $G_{\mathrm{cond}}$ to a negligible value compared to the other (radiative) terms.

In short, by combining Eqs. \ref{Detectivity1} - \ref{eq SW}, we obtain a general expression for the fundamental detectivity limit $D^*_T$ that is valid for all cooling conditions:
\begin{equation} \label{eq Dstar gen}
D^*_T=\frac{\gamma_{\mathrm{abs}}A^{1/2}}{S_P^{1/2}}.
\end{equation}
This expression can be further developed into useful limit cases. The first occurs under active cooling,  if conduction heat transfer is engineered to be negligible ($G_{\mathrm{cond}}\approx0$) together with non-zero opening ($\theta>0$). In this case the $G_{\mathrm{cond}}$ term from \autoref{eq SW} can be neglected, together with the $T_c^5$ term, which will always be smaller than the $T_d^5$ term due to $T_c<T_d<T_b$. The fundamental detectivity limit then reduces to:  
\begin{equation}
D_{T,\mathrm{cooled}}^{*} =
\frac{
\gamma_{\mathrm{abs}}
}{
\left(
\sigma_{SB} k_{B}
\left(
16\,\varepsilon_{T_d}T_{d}^{5}
+ 8\,\varepsilon_{T_b}T_{b}^{5}\sin^{2}\theta
\right)
\right)^{1/2}
} .
\end{equation}
Furthermore, for cooled cases with sufficiently large opening angles $\theta$, temperature fluctuations can become limited by coupling to the radiation background, which is often refereed to as the background-limited performance case (BLIP) \cite{kruse_elements_1962}. In this case, ${D^*_T}$ can be further reduced to  
\begin{equation}
D_{T,{{\mathrm{BLIP}}}}^{*} =
\frac{
\gamma_{{\mathrm{abs}}}
}{
( 8\ \sigma_{SB} k_{B} \varepsilon_{T_b}\,T_{b}^{5}\sin^{2}\theta)^{1/2}
} .
\end{equation}

Conversely, in an uncooled case where $T_c=T_b=T_d=T$, all radiative terms in $S_P$ can be consolidated together, yielding the room temperature detectivity limit:  
\begin{equation}
D_{T,{\mathrm{RT}}}^{*}
= 
\frac{\gamma_{{\mathrm{abs}}}}{(32 \sigma_{SB} k_{B} \varepsilon_{T} T^5+ 4 k_B G_{\mathrm{cond}}T^2/A)^{1/2}}, 
\end{equation}
which reduces to the well known area-independent limit ($\approx10^{10} \:\mathrm{cm^{1/2} Hz^{1/2}W^{-1}} $) \cite{kruse_elements_1962} when $G_{\mathrm{cond}}\approx 0 $.

In \autoref{fig:Detectivity_vs_Tc} (b), we predict the maximum possible detectivity as a function of casing temperature $T_c$ and opening angle $\theta$. We consider a sensor coupled radiatively on both faces, as well as the grey body assumption with the maximum possible emissivity/absorption of a thin metal film absorber ($\varepsilon=\gamma_{\mathrm{abs}}=0.5$)  \cite{luhmann_ultrathin_2020,woltersdorff_uber_1934,hadley_reflection_1947,hilsum_infrared_1955}. In \autoref{fig:Detectivity_vs_Tc}, the cases where $G_{\mathrm{cond}}=0$ are universal limits that are independent of the sensor geometry (area, thickness) or material. To illustrate the effect of non-negligible heat conduction, \autoref{fig:Detectivity_vs_Tc} also includes the example case of $G_{\mathrm{cond}}=8\pi h k$ for a 1 mm square silicon nitride (SiN) membrane of 50 nm thickness, considering $k_{\mathrm{SiN}}=2.7 \; \mathrm{W m^{-1} K^{-1}}$. The background temperature is set to $T_b=300 \:\mathrm{K}$. 

This square SiN membrane sensor geometry is also used throughout the remainder of the manuscript whenever plotting the example behavior of a typical resonator. Other SiN parameters considered throughout the manuscript include specific heat $c_p=700$ J/(kg$\cdot$K), Young's modulus $\mathrm{E} = 300 \:\mathrm{GPa}$, density $\rho=2300$ $\mathrm{kg/m^3}$, thermal conductivity $k=2.7\: \mathrm{W/(m\cdot K)}$, Poisson ratio $\nu=0.28$, thermal expansion coefficient $\alpha_T=2.2\times10^{-6}$ $\mathrm{K}^{-1}$. We also consider that intrinsic quality factor ($Q_{\mathrm{int}}$) normalized by thickness ($q_{\mathrm{int}}=Q_{\mathrm{int}}/h$) has a constant value of $q_{{\mathrm{int}}}=5.7\times 10^{10} \: \mathrm{m}^{-1}$ \cite{villanueva_evidence_2014}. For conciseness, we ignore the temperature dependence of material properties of SiN. We note, however, that analyzing the effect of cooling on material properties (e.g., specific heat, thermal conductivity) would be relevant in future work since they can vary substantially at cryogenic temperature \cite{zink2004specific}.   

An important conclusion of \autoref{fig:Detectivity_vs_Tc} (b) is that cooling does not significantly improve detectivity unless coupling to the room temperature background (at $T_b$) is minimized by reducing the opening angle $\theta$. This can easily be understood by looking at the second term of \autoref{eq SW}, which accounts for thermal fluctuations from the background. This term is unaffected by cooling and only adjustable via $\theta$. We also note that non negligible conduction ($G_{\mathrm{cond}}\neq 0$) typically has a detrimental effect on detectivity. The only exception occurs at very low temperatures, where non negligible conduction helps in reaching lower $T_d$ values, as seen in \autoref{fig:Detectivity_vs_Tc} (a). Plotting detectivity as a function of $T_d$, instead of $T_c$ would indeed suppress all apparent benefit of non-negligible conduction. 

\section{Bandwidth limits} \label{Section bandwidth}
\begin{figure}[!ht]
    \centering
    \includegraphics[width=1.0\linewidth]{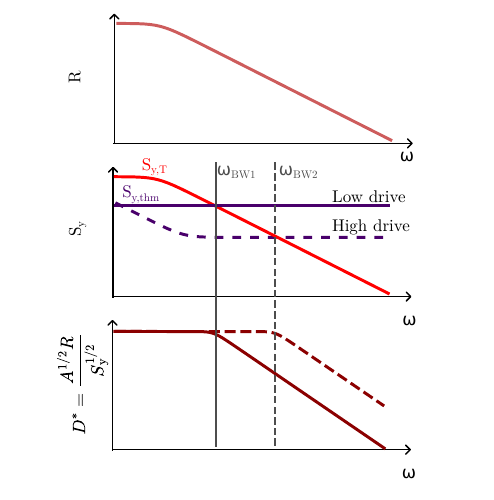}
    \caption{Visual representation of the impact of additive frequency noise ($S_\mathrm{y\mathrm{,thm}}$ in this example) on sensing performances. As shown in Eqs. \ref{Detectivity1} - \ref{eqNEP}, the final detectivity (third panel) is proportional to the ratio of responsivity (first panel) and frequency noise (second panel). The sensor can only reach its maximum detectivity at frequencies where additive frequency noise does not exceed fundamental temperature fluctuation noise $S_\mathrm{y,T}$. This bandwidth is defined by the intersection ($\omega_{\mathrm{BW}}$) of $S_\mathrm{y,\mathrm{thm}}$ and fundamental temperature fluctuations $S_\mathrm{{y,T}}$. Increasing the drive amplitude allows for larger $\omega_{\mathrm{BW}}$, but also causes additional non-linear noise at low frequencies. }
    \label{fig:Detectivity BW schematic}
\end{figure}

The maximum detectivity limits $D_T^*$ derived in the previous section are only reached in cases where the overall resonator frequency fluctuations $S_\mathrm{y}$ are dominated by $S_\mathrm{y,T}$. In other words, $D_T^*$ is reached when additive frequency fluctuations ($S_\mathrm{y\mathrm{,thm}}$ from thermomechanical fluctuations, and $S_\mathrm{y\mathrm{,read}},$ from readout fluctuations) are smaller than fundamental temperature fluctuation noise $S_\mathrm{y,T}$. In practice, this is not a condition satisfied over the entire spectrum of resonator frequency fluctuation, but only over a certain bandwidth $\omega_{\mathrm{BW}}$ which is described by the intersection between the additive noises ($S_\mathrm{y\mathrm{,thm}}$, $S_\mathrm{y\mathrm{,read}}$), and the temperature fluctuation noise spectra $S_\mathrm{y,T}$ \cite{zhang_enhanced_2025}. This is schematized in \autoref{fig:Detectivity BW schematic}, where the sensor reaches maximum $D^{*}$ only over the bandwidth within which thermomechanical fluctuations $S_\mathrm{y\mathrm{,thm}}$ do not exceed fundamental temperature fluctuations $S_\mathrm{y,T}$. Higher drive amplitude increases $\omega_{\mathrm{BW}}$, but there is eventually a limit to this increasing due to the emergence of non-linear thermomechanical frequency noise. 

Previous work have shown that for a linear NMR radiation sensor in a phase lock loop (PLL) tracking scheme, the intersection (in rad/s) between thermomechanical fluctuation noise $S_\mathrm{y\mathrm{,thm}}$ and fundamental temperature noise $S_\mathrm{y,T}$ is given by \cite{zhang_enhanced_2025} 

\begin{equation} \label{Eq:BW thm Chang}
\omega_{\mathrm{BW,thm}} =
\left(
\frac{
S_P\,\alpha^{2}m_{\mathrm{eff}}\omega_{0}^{3}Q_{\mathrm{lin}}a^{2}
}{
2G^{2} k_{B} T_d
}
-1
\right)^{1/2}
\frac{1}{\tau_{\mathrm{th}}} ,
\end{equation}
where ${m_\mathrm{eff}}$ is the effective mass of the NMR, ${\omega_0}$ is the frequency of the driven eigenmode, ${Q_{\mathrm{lin}}}$ is the quality factor of the NMR in the linear regime and $a$ is the driven oscillation amplitude (in m). Likewise the intersection between $S_\mathrm{y\mathrm{,read}}$ and $S_\mathrm{y,T}$ is well approximated by
\begin{equation} \label{Eq:BW read Chang}
\omega_{\mathrm{BW,read}} =
\left( \frac{ S_P \alpha^2 a^2 \omega_0^2}{2G^2 S_\mathrm{x}} \right)^{1/4}
\left( \frac{1}{\tau_{\mathrm{th}}} \right)^{1/2}
\end{equation}
when the displacement readout noise ($S_\mathrm{x}$, in $\mathrm{m^2\:Hz^{-1}}$) is low enough for an intersection to exist between $S_\mathrm{y\mathrm{,read}}$ and $S_\mathrm{y,T}$. Finally, these two intersections can be combined in an effective bandwidth using the phenomenological expression \cite{zhang_enhanced_2025}
\begin{equation}\label{BW}
\omega_{\mathrm{BW}}^{-N} = \omega_{\mathrm{BW,thm}}^{-N}+  \omega_{\mathrm{BW,read}}^{-N},
\end{equation}
which works well in most studied cases for $N\approx3$. 

Although useful, Eqs. \ref{Eq:BW thm Chang} - \ref{Eq:BW read Chang} do not allow prediction of the maximum bandwidth $\omega_{\mathrm{BW, max}}$ as it does not predict the maximum possible driving amplitude ($a$) before nonlinear phenomena deteriorate the frequency stability. A common practice is to drive the resonator at the onset of the emergence of non-linear dynamics \cite{Schmid2023Fundamentals,lifshitz_nonlinear_2008}, with an amplitude $a_\mathrm{c}$. However, recent demonstrations have shown that NMR mechanical performances \cite{manzaneque_resolution_2023,zhang_enhanced_2025} can keep improving well beyond this amplitude despite the emergence of non-linear effects. 

When driving over the critical amplitude, the resonator enters a regime where both stiffness and damping increase with driving amplitude. We refer to these effect as Duffing stiffening \cite{Schmid2023Fundamentals,manzaneque_resolution_2023} and non-linear damping \cite{Schmid2023Fundamentals,catalini_modeling_2021}. As shown in \cite{catalini_modeling_2021}, when lateral displacement is negligible relative to out-of-plane, the Duffing ($\gamma$, in $\mathrm{m}^{-2}$) and non-linear damping coefficients (${\beta}$, in $\mathrm{m}^{-2}$) are correlated as 
\begin{equation}\label{beta}
\beta = \frac{\gamma}{Q_{\mathrm{int}}} ,
\end{equation}
where $Q_{\mathrm{int}}$ is the resonator intrinsic (undiluted) quality factor.

When considering both Duffing non-linearity and non-linear damping, the NMR equation of motion becomes:
\begin{equation} \label{eq Motion}
   \ddot{x}(\hat{t}) + 2\zeta \dot{x}(\hat{t}) + 2\beta x^2(\hat{t}) \dot{x}(\hat{t}) + x(\hat{t}) + \gamma x^{3}(\hat{t}) = \hat{g}(\hat{t}), 
\end{equation}
where $x$ quantifies the displacement of the resonator, $\zeta=1/(2Q_{\mathrm{lin}})$ is the linear damping coefficient, $\hat{t}$ normalizes time $t$ as $\hat{t}=\omega_0 t$ and, similarly, $\hat{g}$ is a normalization of the driving force $g$ by the stiffness of the resonator.

\begin{figure}[!ht]
    \centering
    \includegraphics[width=1.0\linewidth]{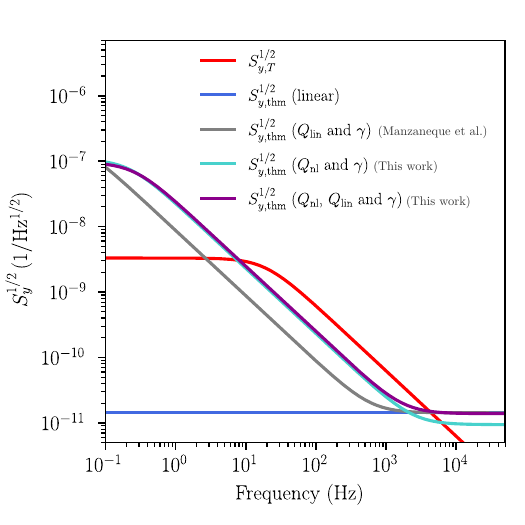}
    \caption{Comparison of different models for thermomechanical frequency noise ($S_\mathrm{y,thm}$) from the literature with the one proposed in this work (magenta), which combines linear damping ($Q_{\mathrm{lin}}$), Duffing nonlinearity ($\gamma$), and nonlinear damping ($Q_{\mathrm{nl}}$). The example resonator considered is described in Section II (1 mm, 50 nm thick SiN), and is driven at its fundamental resonance mode ($n=j=1$), with an amplitude of 1 $\mu$m. Compared with Manzaneque's model \cite{manzaneque_resolution_2023} (in gray, considering $Q_{\mathrm{lin}}$ and $\gamma$), our updated model predicts greater frequency noise at low frequency due to non-linear damping. At such a high drive amplitude, neglecting linear damping (cyan curve, and Eqs. \ref{dev_1} - \ref{dev_last}) results only in a modest underestimation of noise, only at high frequencies. The plot also includes fundamental temperature fluctuations of the same resonator ($S_\mathrm{y,T}$) for comparison purpose.}   
    \label{fig:Chang-Manzaneque Reconciliation}
\end{figure}

To predict the deterioration of frequency stability at high drive amplitude, we rederive Manzaneque's model for non-linear frequency noise \cite{manzaneque_resolution_2023}, but starting from \autoref{eq Motion}, which includes an additional non-linear damping term ($\beta$) not originally included in \cite{manzaneque_resolution_2023}. As in \cite{manzaneque_resolution_2023}, we assume slow dynamic and constant phase shift ($-\pi/2$) in the tracking loop, and we apply the method of averaging. The detailed derivation is given in Appendix \ref{Rederivation Manzaneque}. The resulting thermomechanical frequency noise spectrum of a NMR subjected to nonlinearities is then given by 
\begin{equation} \label{Eq:Sy,thm global}
\begin{aligned} 
S_{\mathrm{y,thm}} ={}&\;
\left(\frac{\omega}{\omega_{0}}\right)^{2}
\Biggl(
\frac{
18\,\gamma^{2}a^{2}k_{B}T_{d}\left(2\zeta + 2a^{2}\beta\right)
}{
\omega^{2}\omega_{0}
\left(
\left(3a^{2}\beta + 4\zeta\right)^{2}
+ \dfrac{16\omega^{2}}{\omega_{0}^{2}}
\right)
m_{\mathrm{eff}}
}
\\[6pt]
&\quad
+
\frac{
\left(a^{2}\beta + 4\zeta\right)^{2}k_{B}T_{d}
}{
2\,\omega_{0}\,\omega^{2}
\left(2\zeta + 2a^{2}\beta\right)
m_{\mathrm{eff}}\,a^{2}
}
\Biggr) ,
\end{aligned}
\end{equation}
where $\omega$ is the fluctuation frequency. As expected, for $\beta=0$, the equation reduces to Manzaneque's original model \cite{manzaneque_resolution_2023}. Conversely, a new limit case arises when non-linear damping dominates over linear damping ($\beta a^2>>\zeta$), which can typically happen for very high driving amplitudes, as in \autoref{fig:Chang-Manzaneque Reconciliation}. In this case we can ignore the linear damping $\zeta$ term and obtain a simplified equation given by 
\begin{equation}\label{dev_1}
S_{\mathrm{y,thm}} =
\frac{
K(\omega^{2} + \chi^{2}\omega_{c}^{2})
}{
\omega_0^{2}(\omega^{2} + \omega_{c}^{2})
} ,
\end{equation}

where 
\begin{equation} \label{Eq:K}
K =
\frac{k_{B}T_d}{8 m_{\mathrm{eff}} Q_{\mathrm{nl}} \omega_{0}a^{2}} ,
\end{equation}

\begin{equation}
\omega_{c} = \frac{3\omega_{0}}{8 Q_{\mathrm{nl}}} ,
\end{equation}

and

\begin{equation}\label{dev_last}
\chi = (64\gamma^{2}a^{4} Q_{\mathrm{nl}}^{2} + 1)^{1/2}.
\end{equation}
$Q_{\mathrm{nl}}$ denotes the quality factor resulting from non-linear damping and is defined by $Q_{\mathrm{nl}}=1/(2\beta a^2)$. For strong noise conversion ($\chi>>1$), three frequency subdomains emerge where frequency noise can be described by the asymptotes  given by 
\begin{equation}
S_\mathrm{y,thm}(\omega \ll \omega_{c}) =
\left( \frac{\omega}{\omega_{0}} \right)^{2}K\,\chi^{2}\,\frac{1}{\omega^{2}} .
\end{equation}

\begin{equation}
S_\mathrm{y,thm}(\omega_{c} \ll \omega \ll \chi\omega_{c}) = \left( \frac{\omega}{\omega_{0}} \right)^{2}
K\,\chi^{2}\omega_{c}^{2}\,\frac{1}{\omega^{4}} .
\end{equation}

\begin{equation}
S_\mathrm{y,thm}(\omega \gg \chi\omega_{c}) = \left( \frac{\omega}{\omega_{0}} \right)^{2}
K\,\frac{1}{\omega^{2}} .
\end{equation}
A similar asymptotic form accounting for both $Q_{\mathrm{lin}}$ and $Q_{\mathrm{nl}}$ is given in Appendix \ref{Rederivation Manzaneque}.

In \autoref{fig:Chang-Manzaneque Reconciliation}, we note that our proposed model for $S_\mathrm{y,thm}$ (in purple, considering $Q_{\mathrm{nl}}$, $Q_{\mathrm{lin}}$ and $\gamma$) predicts more fluctuations than Manzaneque's original model \cite{manzaneque_resolution_2023}, which did not account for non-linear damping  (in gray, considering $Q_{\mathrm{lin}}$ and $\gamma$). Excess frequency fluctuations are especially pronounced at low frequency, i.e., in the range $\omega<\omega_\mathrm{c_{high}}$, where $\omega_\mathrm{c_{high}}$ is defined in \autoref{wchigh} ($\approx 1000\:\mathrm{Hz}$ in this example). In contrast, in the range defined by $\omega>\omega_\mathrm{c_{high}}$, nonlinear phenomena are less pronounced, and both models eventually converge to the linear expression of thermomechanical frequency noise of Refs. \cite{Schmid2023Fundamentals,zhang_enhanced_2025,demir_understanding_2021} (labeled ``$S_\mathrm{y,thm}$ (linear)", in blue). For the same reason, neglecting linear damping altogether (cyan curve considering $Q_{\mathrm{nl}}$ and $\gamma$), yields good results everywhere except in the range $\omega>\omega_\mathrm{c_{high}}$. Despite this slight discrepancy, neglecting linear damping remains useful for simplified noise modeling  (i.e.,  Eqs. \ref{dev_1} - \ref{dev_last}). Furthermore, the discrepancy remains somewhat modest and it would grow even smaller for resonators of higher $Q_\mathrm{lin}$ than the example square membrane resonator considered here (e.g., in soft-clamped resonators). 

It should also be noted that Manzaneque's \cite{manzaneque_resolution_2023} original model, and the expanded version proposed herein consider an infinite bandwidth of the frequency tracking loop, while the linear models in \cite{Schmid2023Fundamentals,zhang_enhanced_2025,demir_understanding_2021} can account for different bandwidths of frequency tracking (e.g., different PLL parameters). To compare all models together in \autoref{fig:Chang-Manzaneque Reconciliation}, we set an infinitely large tracking bandwidth for all models. It has been shown in \cite{zhang_enhanced_2025}, that the PLL tracking parameters have no fundamental effect on the final bandwidth $\omega_{\mathrm{BW}}$, making the infinite bandwidth assumption sufficient for the current work. 


\begin{figure}[!ht]
    \centering
    \includegraphics[width=1.0\linewidth]{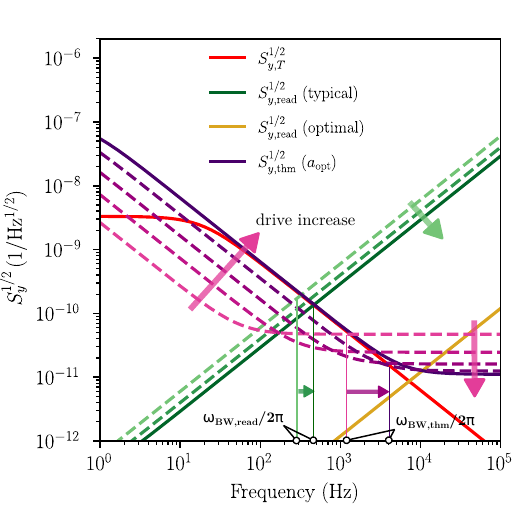}
    \caption{Increasing the oscillation amplitude $a$ until the optimal value $a_\mathrm{opt}$ that maximizes the thermomechanical bandwidth limit $\omega_\mathrm{BW,thm}$. Increasing $a$ also improves readout noise (green traces) which, in this example, is calculated for a typical displacement readout value $S_\mathrm{x}\approx50\: \mathrm{fm\:Hz^{-1/2}}$). In this particular example of a square SiN membrane at room temperature (see resonator properties in Section II), the overall bandwidth remains readout limited, unless the readout noise is significantly improved down to the yellow ``$S_\mathrm{y,read}$ (optimal)" trace.} 
    \label{fig:RT sensor fluctuations}
\end{figure}

With this new model for linear and non-linear thermomechanical frequency noise, we can finally find the optimal drive amplitude $a_{\mathrm{opt}}$ that maximizes the intersection frequencies $\omega_{\mathrm{BW}}$, while maintaining nonlinear thermomechanical noise $S_\mathrm{y,thm}$ below fundamental temperature fluctuation noise $S_\mathrm{y,T}$. As shown in \autoref{fig:RT sensor fluctuations}, by increasing the drive amplitude until $S_\mathrm{y,thm}$ matches $S_\mathrm{y,T}$, we maximize the value of $\omega_{\mathrm{BW,thm}}$ by pushing the linear part of $S_\mathrm{y,thm}$ to values as low as possible. Therefore, the optimal performance scenario for NMR radiation sensor is the overlay of temperature and thermomechanical noise spectral densities. This overlay can only occur between the thermal cutoff frequency ($\omega_{\mathrm{th}}=1/\tau_{\mathrm{th}}$) and $\chi \omega_c$ (see \autoref{fig:RT sensor fluctuations}) where frequency noise equations respectively reduce to:
\begin{equation}\label{ThermalSlope}
S_\mathrm{y,T}(\omega) \approx \frac{S_P}{G^2} \, \alpha^2 \, 
\frac{1}{\omega^2\tau_{\mathrm{th}}^2}
\end{equation}
and
\begin{equation}\label{thermomechanical slope}
S_\mathrm{y,{\mathrm{thm}}}(\omega) \approx \frac{9}{8}\frac{k_BT_da^2\gamma^2}{Q_{\mathrm{nl}} m_{\mathrm{eff}}\omega_0\omega^2}.
\end{equation}

We therefore define the optimal drive amplitude $a_\mathrm{opt}$ by the one that results in Eqs. \ref{ThermalSlope} and \ref{thermomechanical slope} to be equal. The optimal driving amplitude in the context of NMR-based radiation sensing therefore reduces to:
\begin{equation} \label{Eq:aopt}
a_{\mathrm{opt}}
=
\left(
\frac{4}{9}
\frac{
S_P\,\alpha^{2} m_{\mathrm{eff}}\omega_{0}
}{
G^{2} \tau_{\mathrm{th}}^{2}
k_{B}T_{d}\gamma^{2}\beta
}
\right)^{1/4} .
\end{equation} 
Interestingly, this optimal driving amplitude for an NMR radiation sensor can be well above the traditional linear regime limit \cite{Schmid2023Fundamentals} given by
\begin{equation}
a_c=\frac{4}{3^{3/4}}\left(\frac{\zeta}{\gamma}\right)^{1/2},  
\end{equation} 
thus contradicting the common assumption that maximum performances are reached at the onset of non-linearity. Strikingly, for our example resonator membrane (described in Section II), the optimal drive amplitude  $(a_{\mathrm{opt}}=1.6\times10^{-6})$ is two orders of magnitude higher than the classical limit ($a_{\mathrm{c}}=1.2\times10^{-8}$ m). This surprising result is consistent with experimental results in \cite{zhang_enhanced_2025}, where resonators were sucessfully operated beyond the classical limit by a comparable amount. 

We note that, in practice, an exact overlay of the additive noise over the fundamental temperature noise (as in \autoref{fig:RT sensor fluctuations}) is not exactly desirable as this makes $S_\mathrm{y,thm}$ non-negligible. To be more precise, the optimal amplitude yielding the best sensing performance resides just at the onset of this overlay case. Likewise, as shown, in \autoref{fig:RT sensor fluctuations} an exact overlay leads to excess noise below the thermal corner frequency $\tau_{\mathrm{th}}^{-1}$. The actual bandwidth over which thermal fluctuations dominate could therefore be corrected as $\omega_{\mathrm{BW,eff}}=\omega_{\mathrm{BW}}-\tau_{\mathrm{th}}^{-1}$. These correction are typically small, such that we omit them for conciseness in the remainder of this work.

When driving at the optimal amplitude $a_\mathrm{opt}$, the thermomechanical bandwidth limit can be found by substituting $a_\mathrm{opt}$ in \autoref{Eq:BW thm Chang}
\begin{equation}  \label{Eq BW thm}
\omega_{\mathrm{BW,thm}}
=\left(
\frac{
\alpha^{6}\,
\omega_{0}^{7}\,
S_P^{3}\,
m_{\mathrm{eff}}^{3}\,
Q_{\mathrm{lin}}^{2}
}{
9G^{6}\,
\tau_{\mathrm{th}}^{6}\,
k_{B}^{3}\,
T_d^{3}\,
\gamma^{2}\,
\beta}\right)^{1/4}.
\end{equation}
Likewise, the read-out limit bandwidth is found by substituting $a_\mathrm{opt}$ in \autoref{Eq:BW read Chang}
\begin{equation} \label{Eq BW read}
\omega_{\mathrm{BW,read}}
=
\left(\frac{
\alpha^{6}\,
S_P^{3}\,
m_{\mathrm{eff}}\,
\omega_{0}^{5}
}{
9G^{6}\,
\tau_{\mathrm{th}}^{6}\,
S_\mathrm{x}^{2}\,
k_{B}T_d\,
\gamma^{2}\beta\,
}\right)^{1/8}.
\end{equation}
Note that \autoref{Eq BW thm} assumes that frequency noise is limited by linear damping in the range $\omega>\omega_\mathrm{c_{high}}$, as discussed in the context of \autoref{fig:Chang-Manzaneque Reconciliation}. Negligible linear damping (i.e., very high $Q_\mathrm{lin}$) would instead result in the ultimate bandwidth limit $\omega_\mathrm{UBW}$ discussed later in this section. 

Building on Eqs. \ref{Eq BW thm} - \ref{Eq BW read} we can derive the maximum acceptable noise $S_\mathrm{x}$ (in $\mathrm{m^2\: Hz^{-1}}$) of the NMR displacement readout, such that readout noise does limit the sensor bandwidth. To respect this condition, $S_\mathrm{x}$ should be minimized such that $\omega_\mathrm{BW,read}>\omega_\mathrm{BW,thm}$. By inspection of Eqs. \ref{Eq BW thm} - \ref{Eq BW read}, this is possible for: 
\begin{equation}
    S_\mathrm{x}<\frac{2G^2\tau_{\mathrm{th}}^2k_B^2T_d^2}{S_P\alpha^2a_{\mathrm{opt}}^2\omega_0^4m_{\mathrm{eff}}^2Q_{\mathrm{lin}}^2},
\end{equation}
which can be further developed by substituting the value of $a_\mathrm{opt}$ from \autoref{Eq:aopt}:
\begin{equation}\label{Eq:Sxx}
S_\mathrm{x}<\left(\frac{
9G^{6}\,\tau_{\mathrm{th}}^{6}\,
k_{B}^{5}\,T_d^{5}\,
\gamma^{2}\beta\,
}{
Q_{\mathrm{lin}}^{4}\,
S_{P}^{3}\,
\alpha^{6}\,
m_{\mathrm{eff}}^{5}\,
\omega_{0}^{9}
}
\right)^{1/2}.
\end{equation}

This requirement on the readout displacement noise is illustrated in \autoref{fig:RT sensor fluctuations}, where we include the readout frequency fluctuations respecting \autoref{Eq:Sxx} ($S_\mathrm{y,read}$ ``optimal") alongside the readout noise ($S_\mathrm{y,read}$ ``typical") resulting from a typical $S_\mathrm{x}$ experimental value (i.e., $S_\mathrm{x}^{1/2}\approx 50 \; \mathrm{fm \: Hz^{-1/2}}$ \cite{rugar_improved_1989,zhang_enhanced_2025}). In this particular example, the typical experimental readout noise is clearly the main performance limiter, deteriorating the overall bandwidth by one order of magnitude compared with the optimal case. Nevertheless, sufficiently low interferometer static noises have been achieved in previous work \cite{Underwood2015} and are therefore realistic in future work. 

\begin{figure}[!ht]
    \centering
    \includegraphics[width=1.0\linewidth]{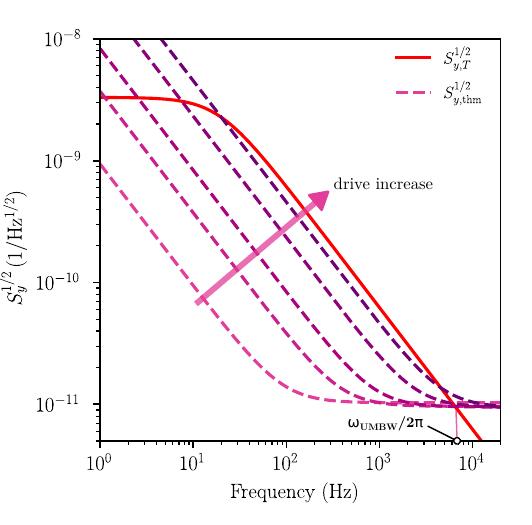}
    \caption{In the limit of very high linear quality factor $Q_\mathrm{lin}$, an ultimate bandwidth limit $\omega_\mathrm{UMBW}$ emerges and is limited by nonlinear damping. The figure is simulated for the room temperature square resonator parameters as given in Section II, but by setting clamping losses to zero to approximate a soft-clamped resonator. As opposed to the linear damping case of \autoref{fig:RT sensor fluctuations}, this limit bandwidth is independent of the drive amplitude $a$, as long as $a>a_\mathrm{UBW}$. } 
    \label{fig:SC Sensor Fluctuations}
\end{figure}

Lastly, an interesting other limit case arises in the limit of very high quality factor $Q_\mathrm{lin}$. In this case, the high frequency part of the thermomechanical frequency fluctuations spectrum are quickly limited by non-linear damping and therefore become amplitude-independent (see \autoref{fig:SC Sensor Fluctuations}). This can be understood by noting that the product of $Q_\mathrm{nl}$ and $a^2$ in \autoref{Eq:K} is constant and equals to $(2\beta)^{-1}$. 

This hard limit on $\omega_\mathrm{BW}$ corresponds to the ultimate theoretical limit of thermomechanical noise improvement that can be achieved through increasing the drive amplitude. This theoretical ultimate maximum bandwidth ($\omega_{\mathrm{UMBW}}$) is given by
\begin{equation}
    \omega_{\mathrm{UBW}}=\left(\frac{4\,S_P\,\alpha^{2}\,m_{\mathrm{eff}}\,\omega_{0}^{3}
    }{
    G^{2}\,\tau_{\mathrm{th}}^{2}\,k_{B}\,T_d\,\beta
    }
    \right)^{1/2}
\end{equation}
and is reached for driving amplitudes larger than
\begin{equation}
    a_{\mathrm{UBW}}=4\left(\frac{\zeta}{ \beta }\right)^{1/2}.
\end{equation}
Although an interesting case, reaching the $\omega_{\mathrm{UMBW}}$ is challenging in the context of NMR radiation sensors, because thermomechanical fluctuations from linear damping and readout noise typically limit the bandwidth at high fluctuation frequencies. Nevertheless, high $Q_\mathrm{lin}$ resonators based on soft clamping of low order modes, such as \cite{shin_spiderweb_2022,bereyhi_hierarchical_2022}, are a promising avenue, if they can be coupled with very high performance displacement readout. 

\section{Detectivity $\times$ Bandwidth limits}

Having derived closed form expressions for NMR sensor performance metric that are the detectivity (Sec. II) and bandwidth (Sec. III), we can now provide a universal figure of merit (FOM) as a weighted product of $D^*$ and $\omega_\mathrm{BW}/2\pi$ (where the $2\pi$ term sets all frequency units to Hz). Such an FOM is needed for optimization of resonator parameters since some changes can be beneficial for one performance metric while being detrimental to the other, making optimization difficult without a single global figure of merit. While many weighing factors are possible, we find that   
\begin{equation}
\mathrm{FOM} = D^{*\:3/2} \cdot \frac{\omega_{\mathrm{BW}}}{2\pi},
\end{equation}
is the most useful for two main reasons. Firstly, by inspection of \autoref{eq Dstar gen} and \autoref{Eq BW thm}, we find that this weighing makes the FOM independent of the relatively cumbersome $S_P$ term when the sensor bandwidth is limited by thermomechanical fluctuations. This yields a remarkably simple, single-term FOM as the thermomechanical performance limit of NMR radiation sensors: 
\begin{equation}\label{FOM thm}
\mathrm{FOM_{thm}}
=
\left(\frac{
\varepsilon^{6}A^{3}\alpha^{6}\,
\omega_{0}^{7}\,
m_{\mathrm{eff}}^{3}\,
Q_{\mathrm{lin}}^{2}
}{
144\pi^4\,
G^{6}\,
\tau_{\mathrm{th}}^{6}\,
k_{B}^{3}\,
T_d^{3}\,
\gamma^{2}\beta\,
}\right)^{1/4}
\end{equation}
Secondly, we find that over-weighing of detectivity relative to bandwidth conveniently prevents optimization results from leaning towards high bandwidth, low detectivity configurations. Such a result would be less relevant in practical settings, since the high detectivity of NMR-based sensors is typically their main distinctive feature compared to other technologies. For this reason, we keep the same weighing factor when performances are limited by readout noise, even though the $S_P$ term does not cancel in this particular case. By combining \autoref{eq Dstar gen} and \autoref{Eq BW read}, we obtain the following FOM when the sensor bandwidth is readout-limited: 
\begin{equation}\label{FOM read}
\mathrm{FOM_{read}}
=
\left(\frac{
A^{6}\varepsilon^{12}
\alpha^{6}\,
m_{\mathrm{eff}}\,
\omega_{0}^{5}
}{
2304\pi^4\,
S_P^{3}\,
G^{6}\,
\tau_{\mathrm{th}}^{6}\,
S_{x}^{2}\,
k_{B}T_d\,
\gamma^{2}\beta\,
}
\right)^{1/8}
\end{equation}

\section{Dimensional Analysis}
While the figure of merits in Sec. IV are generally applicable to any NMR-based radiation sensor, it is insightful to derive their value for the simple case of a square drum resonator, for which analytical expressions exist for all terms in \autoref{FOM thm} and \autoref{FOM read}. Doing so can provide key insights to generally unanswered questions on, for example, optimal resonator dimensions, tensile stress, or the importance of quality factor.

For this analysis, we consider the eigenfrequencies (in rad/s) of a square Duffing membrane resonator, which are given by
\begin{equation}\label{eigenfrequencies}
\omega_{0}=\frac{\pi\,(n^{2} + j^{2})^{1/2}}{L}\left(\frac{\sigma}{\rho}\right)^{1/2}
\left( 1 + \frac{3}{8}\,\gamma\,a^{2} \right),
\end{equation}
where $n$ and $j$ are the mode order of the resonator and $L$ is the side length. For most square resonators, the Duffing frequency shift of \autoref{eigenfrequencies} is very small ($\frac{3}{8}\,\gamma\,a^{2}<<1$), therefore it is neglected in the expression of $\omega_{0}$ in the following derivations. 

The Duffing coefficient ($\gamma$) of a square drum resonator is also analytically derivable \cite{lu_nonlinear_2020,Schmid2023Fundamentals} and is given by 
\begin{equation}\label{Duffing coefficient}
\gamma =
\frac{3}{16}\frac{E}{\sigma}
\frac{\pi^{2}(n^{4}+j^{4})}{L^{2}(n^{2}+j^{2})}
\end{equation}

Lastly, we can also extract both the linear ($Q_\mathrm{lin}$) and non-linear ($Q_\mathrm{nl}$) components of the quality factor of a square drum resonator through simple expressions: 
\begin{equation}\label{Q_lin}
\begin{aligned}
Q_{\mathrm{lin}}
&=\Biggr(
\underbrace{
\frac{\pi^{2}(n^{2}+j^{2})}{12}\,
\frac{E}{\sigma}
\left(\frac{h}{L}\right)^{2}
}_{\text{sine shape losses}}
\\[6pt]
&\quad+\;
\underbrace{
\left(\frac{E}{3\sigma}\right)^{1/2}
\left(\frac{h}{L}\right)
}_{\text{edge losses}}
\Biggr)^{-1}
q_{\mathrm{int}}\,h .
\end{aligned}
\end{equation}
and
\begin{equation}\label{Qnl}
    Q_{\mathrm{nl}}=\left(
\frac{3 \pi^{2} (n^{4}+j^4)}{8(n^2+j^2)}\,
\frac{E}{\sigma}\left(\frac{a}{L}\right)^{2}
\right)^{-1}q_{\mathrm{int}}h
\end{equation}
where $q_\mathrm{int}=Q_\mathrm{int}/h$ is the intrinsic quality factor ($Q_\mathrm{int}$) normalized by the resonator thickness $h$. As shown in \cite{villanueva_evidence_2014}, $q_\mathrm{int}$ can typically be considered constant in thin resonators where mechanical losses are surface-limited. 

In \autoref{Qnl}, we derive $Q_\mathrm{nl}$ from the Duffing parameter of \autoref{Duffing coefficient}, combined with \autoref{beta}. Some sources \cite{Schmid2023Fundamentals} raise doubt on the validity of this method due to non negligible lateral displacement in the resonator at high drive amplitude. We therefore perform finite element simulations (See Appendix \ref{FEM}) to confirm that the assumption of negligible in-plane displacement is valid for the typical drive amplitude and membrane geometry in the present work.

The equations above enable analytical prediction of FOMs for both symmetric ($n=j$) and asymmetric ($n\neq j$) modes. However, for further derivations, only symmetrical modes are considered since (1) symmetric modes typically have the best mechanical properties (i.e., lowest dissipation), and (2) setting $n=j$ allows for substantial algebraic simplifications (for example in \autoref{Duffing coefficient}). 

The thermal coefficient of fractional frequency shift of a square membrane under tensile stress can also be analytically approximated as 
\begin{equation}
    \alpha\approx\frac{E\alpha_T}{2\sigma(1-\nu)},
\end{equation}
The thermal conductance $G$ can be simply expressed as a combination of radiative (left hand side) and conductive (right hand side) as   
\begin{equation}
    G=8\varepsilon_{T_d}\sigma_{\mathrm{SB}}L^2T_d^3+8\pi h k.
\end{equation}
The heat capacity ($C$) and effective mass ($m_{\mathrm{eff}}$) of a membrane can be expressed as  
\begin{equation}
    C=\rho L^2hc_p,
\end{equation}
and
\begin{equation}\label{effective_mass}
    m_{\mathrm{eff}}=\frac{\rho L^2h}{4}.
\end{equation}

With these analytical expressions, FOMs can take very simple forms that depend only on resonator geometry and material properties. When the edge loss term in \autoref{Q_lin} dominates  (i.e., in the commonly called ``hard clamped" scenario) the FOM reduces to 
\begin{equation}\label{FOM thm square membrane}
\begin{aligned}
\mathrm{FOM}_\mathrm{thm,h}
= 0.130\,
\frac{
\alpha_{T}^{3/2}L^{1/4}E^{1/2}\,
n^{1/4}\sigma^{3/8}\,q_{\mathrm{int}}^{3/4}\,
\varepsilon^{3/2}
}{
(1-\nu)^{3/2}C_p^{3/2}h^{1/2}\,
\rho^{13/8}\,T_d^{3/4}k_{B}^{3/4}
},
\end{aligned}
\end{equation}
where we have also assumed $G\approx G_\mathrm{rad}$, as well as the gray body assumption on emissivity values. These assumptions hold for the remainder of the limit case equations given in this section. Interestingly, this FOM is independent of the casing geometry (i.e., of $\theta$) due to our choice of weighing factors in the FOM (see Sec. IV).  
\autoref{FOM thm square membrane} indicates that higher order modes are desirable to increase the FOM. Using high order modes will eventually make the sine shape term dominant in \autoref{Q_lin}. In this case, the ``soft clamped" FOM becomes: 
\begin{equation}\label{FOM thm soft clamp}
\begin{aligned}
\mathrm{FOM}_\mathrm{thm,s}
=0.077
\frac{
\alpha_T^{3/2}\,
L^{3/4}\,
E^{1/4}\,
\sigma^{5/8}\,
q_{\mathrm{int}}^{3/4}\,
\varepsilon^{3/2}\,
}{
(1-\nu)^{3/2}\,
C_p^{3/2}\,
\rho^{13/8}\,
n^{3/4}\,
T_d^{3/4}\,
h\,
k_B^{3/4}
}
\end{aligned}
\end{equation}
However, \autoref{FOM thm square membrane} also shows that the FOM improves very slowly with mode order as $n^{1/4}$, while high mode order also rapidly increases the requirement on readout noise $S_\mathrm{x}$ (see \autoref{Sxx square membrane} below). In practice, the hard clamped case (low mode order) is therefore more likely to occur.

The drive amplitude needed for reaching the thermomechanical FOM of \autoref{FOM thm square membrane} is not as simply expressed since it is not independent of $S_P$. We can nevertheless derive simple expressions in certain limit cases, i.e., when the NMR sensor is either background limited (BLIP) or at room temperature, and with $G\approx G_\mathrm{rad}$:
\begin{equation}
\begin{aligned}
a_{\mathrm{opt}}\label{optimal amplitude square membrane}
= 0.628
\frac{
\sigma_{\mathrm{sb}}^{1/4}\,\varepsilon^{1/4}\,
F^{1/2}\,\alpha_{T}^{1/2}\,
\sigma^{3/8}\,q_{\mathrm{int}}^{1/4}
}{
n^{5/4}\,\rho^{3/8}\,C_p^{1/2}\,
E^{1/4}\,T_{d}^{1/4}\,(1-\nu)^{1/2}
}
\\[4pt]
\times\;
T_{b}^{5/4}\,L^{5/4} ,
\end{aligned}
\end{equation}
where $F=\sin \theta$ in the BLIP case, and $F=2$ in the room temperature case. In the latter, all $T$ values must also be set to room temperature.

Conversely, the bandwidth in the thermomechanical-limited case can be calculated using

\begin{equation}
\begin{split}
\omega_{\mathrm{BW,thm}}
= 3.872\,
\frac{
\alpha_{T}^{3/2}\,L^{1/4}\,
\sigma_{\mathrm{sb}}^{3/4}\,\varepsilon^{3/4}\,
F^{3/2}\!\,T_{b}^{15/4}\,
}{
C_p^{3/2}(1-\nu)^{3/2}
}
\\[4pt]
\times
\frac{
E^{1/2}\,n^{1/4}\,\sigma^{3/8}\,
\mathrm{q_{int}}^{3/4}
}{
\rho^{13/8}\,T^{3/4}\,h^{1/2}
},
\end{split}
\end{equation}
which can also be adapted to the BLIP or room temperature using the same $F$ values. 

As stated previously, reaching this thermomechanical performance limit is only possible if the readout noise $S_\mathrm{x}$ is sufficiently small, which amounts to respecting the condition:
\begin{equation}\label{Sxx square membrane}
\begin{aligned}
S_\mathrm{x} <
0.035\,
\frac{
k_{B}\,C_p^{3}(1-\nu)^{3}
T_{d}^{5/2}\rho^{11/4}
}{
\alpha_{T}^{3}
F^{3}\,T_{b}^{15/2}
n^{3/2}\sigma^{7/4}
q_{\mathrm{int}}^{5/2}E^{1/2}
}
\\[4pt]
\times\;
\frac{1}{L^{5/2}\sigma_{\mathrm{SB}}^{3/2}\varepsilon^{3/2}} .
\end{aligned}
\end{equation}

If this condition is not met, the FOM becomes limited not by thermomechanical noise, but by readout noise:
\begin{equation}\label{FOM read square membrane}
\begin{split}
\mathrm{FOM_{read}}
= 0.056\,
\frac{
\alpha_{T}^{3/4}\,E^{3/8}\,q_{\mathrm{int}}^{1/8}\,
\varepsilon^{9/8}\,\!
}{
F^{3/4}C_p^{3/4}(1-\nu)^{3/4}S_\mathrm{x}^{1/4}k_{B}^{1/2}
}
\\[4pt]
\times
\frac{
1
}{
T_{b}^{15/8}\,T_{d}^{1/8}\,n^{1/8}\,\rho^{15/16}\,
\sigma_{\mathrm{sb}}^{3/8}\,
\sigma^{1/16}\,L^{3/8}h^{1/2}
}
\end{split}
\end{equation}

Likewise, the bandwidth in the readout limited case is then given by
\begin{equation}
\begin{split}
\omega_{\mathrm{BW,read}}
= 1.671\,
\frac{
k_{B}^{1/4}\,T_{b}^{15/8}\,
F^{3/4}\!\,\alpha_{T}^{3/4}\,
E^{3/8}\,\sigma_{\mathrm{sb}}^{3/8}\,
}{
C_p^{3/4}(1-\nu)^{3/4}\,
S_\mathrm{x}^{1/4}\,T_{d}^{1/8}\,n^{1/8}
}
\\[4pt]
\times
\frac{
\varepsilon^{3/8}\,\mathrm{q_{int}}^{1/8}
}{
\rho^{15/16}\,\sigma^{1/16}\,
L^{3/8}\,h^{1/2}
},
\end{split}
\end{equation}
which is achieved using the same drive amplitude as given in \autoref{optimal amplitude square membrane} for the thermomechanical limit. 

\begin{figure}[!ht]
    \centering
    \includegraphics[width=1.0\linewidth]{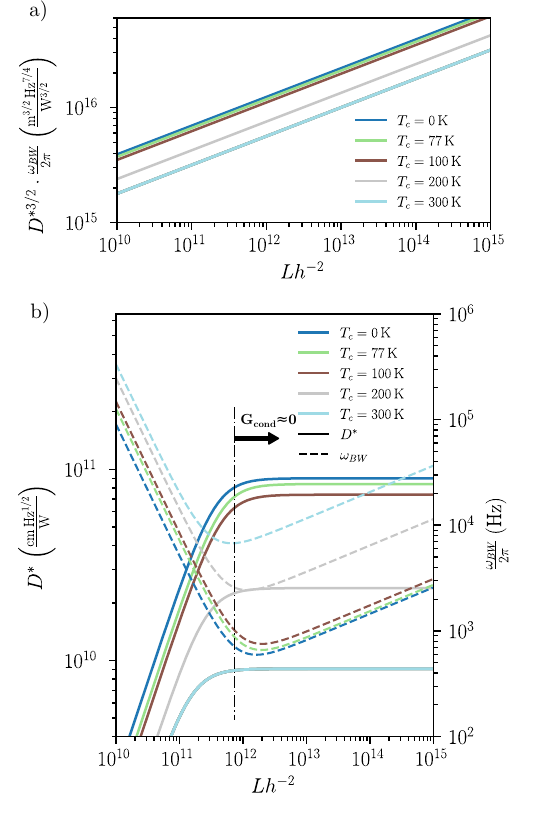}
    \caption{Theoretical maximum figure of merit (a) and associated effective bandwidth and detectivity (b) of a square NMR membrane radiation sensor limited by thermomechanical fluctuations. The background temperature is set at 300 K. For panel (b), a casing opening of 10° is considered, and a thickness of 50 nm is considered to evaluate $G_\mathrm{cond}$. The plots are otherwise valid for any thickness in the range of $G_\mathrm{cond}\approx0$. Likewise, panel (a) is valid for any resonator thickness and opening angle $\theta$. Performances are found to scale as $Lh^{-2}$, meaning that thermomechanically-limited sensors should have an as high aspect ratio as possible}  
    \label{fig:Main results}
\end{figure}

\begin{figure}[!ht]
    \centering
    \includegraphics[width=1.0\linewidth]{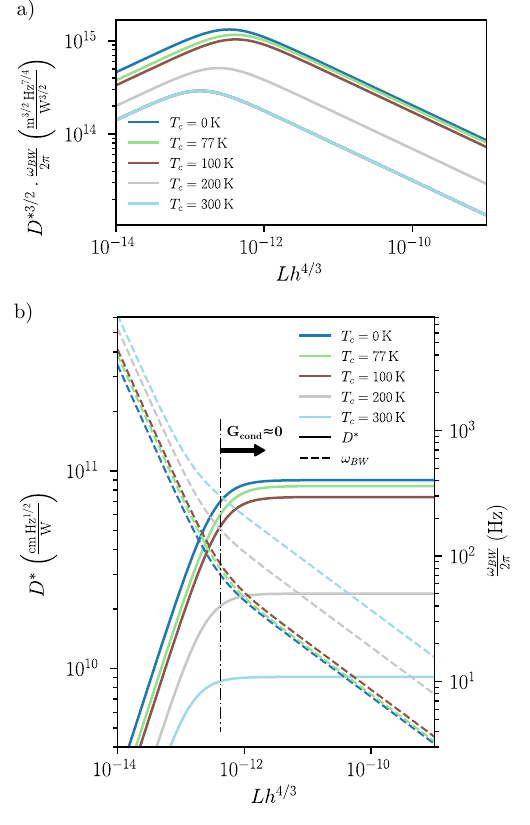}
    \caption{Theoretical maximum figure of merit (a) and associated effective bandwidth and detectivity (b) of a NMR membrane radiation sensor limited by displacement readout noise, assuming $S_x^{-1/2}\approx50\; \mathrm{fm\: Hz^{-1/2}}$. A casing opening of 10° is considered, and a thickness of 50 nm is considered to evaluate $G_\mathrm{cond}$. The plots are otherwise valid for any resonator thickness in the range of $G_\mathrm{cond}\approx0$. For readout limited sensors $h$ should be small, and $L$ should just large enough to make $G_{\mathrm{cond}}$ negligible.}
    \label{fig:ROL results}
\end{figure}

Theoretical FOMs of square drum resonator radiation sensor operating at optimal amplitude for a variety of conditions and parameters are shown in \autoref{fig:Main results} and \autoref{fig:ROL results}. Those figure use the complete relations of Eqs. \ref{FOM thm} - \ref{FOM read} with the full membrane parameters given in Eqs. \ref{eigenfrequencies} - \ref{effective_mass}, and the optimal drive amplitude given by \autoref{Eq:aopt}. The membrane fundamental eigenmode (n=1) is considered. 

We find that optimal resonator parameters are drastically different depending on whether the performances are thermomechanical (\autoref{fig:Main results}) or readout-limited (\autoref{fig:ROL results}). In the thermomechanically-limited case of \autoref{fig:Main results}, the FOM scales linearly with the $L/h^2$ ratio, meaning that very high aspect ratio resonators generally maximize performances. Eqs. \ref{FOM thm square membrane} - \ref{FOM thm soft clamp} also indicate that high stress $\sigma$ and high $q_\mathrm{int}$ are beneficial. We can therefore generally expect that high stress, high $Q$ devices of very high aspect ratio will benefit radiation sensing performances in thermomechanically-limited sensors.

 
Optimal resonator parameters are strikingly different for readout-limited sensors. In this case, the FOM scale as $L^{-1}h^{-4/3}$ (see \autoref{FOM read square membrane}). Thin membranes are therefore still effective at increasing the FOM, but the dimensions $L$ should also be minimized as much as possible, but not so much that $G_{cond}$ becomes significant. This gives rise, in \autoref{fig:ROL results}, to an optimal $L$ value that maximizes the FOM for any given $h$. Likewise, in contrast to the thermomecanical-limited case, low tensile stress $\sigma$ is beneficial to the FOM. $q_\mathrm{int}$ remains beneficial, but to a much lower extend due to the power 1/8 in \autoref{FOM read square membrane}. Readout-limited sensors should therefore be designed to be very thin, with a corresponding optimal dimension $L$. Mechanical quality factor should be a secondary consideration in these cases, as long as $Q$ is not so low that thermomechanical fluctuations start dominating. 

The benefits of cooling are also highlighted. For either FOM, lower casing temperature reduces the effective bandwidth $\omega_\mathrm{BW}$ through lowered thermal time response $\tau_{th}$ (i.e., lowered $G_\mathrm{rad}$). This degradation is however trumped by the benefit of cooling on the detectivity, making cooling generally beneficial to FOMs. It should also be noted that the impact of cooling temperature is highly dependent on aperture opening angle and background temperature, and that only a single case is presented in \autoref{fig:ROL results}. Likewise, future work should investigate more deeply the effect of cooling on the resonator material properties. 



\section{Conclusion}
We developed a new set of analytical models to predict the detectivity of NMR radiation sensors under various cooling conditions. The model emphasizes the critical roles of enclosure temperature and opening angle in optimizing sensor performance. We also improved the prediction of the maximum effective bandwidth of NMR detectors by defining an optimal resonator drive amplitude. Contrary to the commonly accepted assumption, this optimal amplitude is well beyond the linear regime limit. Finally, we unified both models into a single figure of merit that balances the importance of effective bandwidth and detectivity. This figure of merit enable us to draw important conclusions on the ideal geometry of sensors, which can vary dramatically depending on if a sensor is limited by thermomechnical or readout noise. We therefore expect this figure merit to become a key tool for the design of high bandwidth nanomechanical radiation sensors in future work.

\appendix
\section{Heat transfer balance at steady-state for enclosed resonator} \label{Steady state eq}
By assuming that the sensor is in steady state in the casing presented in \autoref{fig:Visualisation}, energy conservations imposes the following conditions, which can be solved to determine the detector temperature:
\begin{multline}
G_{\mathrm{cond}}(T_d-T_c)+\sigma A(\epsilon_{T_d}T_d^4-\epsilon_{T_c}T_c^4)(1+\cos^2\theta)\\+\sigma A(\epsilon_{T_d}T_d^4-\epsilon_{T_b}T_b^4)\sin^2\theta=0.
\end{multline}
This equation can be used in lieu of the one in the main text to determine $T_d$ when conduction heat transport is non negligible, or when the gray body approximation does not hold. The equation is valid for a sensor coupled on both faces (i.e., a freestanding film). For a single-sided coupled sensor, the $(1+\cos^2\theta)$ term simply becomes $\cos^2\theta$.

\section{Linearized model of frequency fluctuations of a resonator under PLL based tracking scheme}\label{Rederivation Manzaneque}
This appendix essentially reproduces the derivations given in \cite{manzaneque_resolution_2023}, but by adding an additional $\beta$ term to account for non linear damping at high drive amplitudes. Readers should consult the original work for more details on the derivation. 

Assuming slow variation of amplitude and phase compared to resonance frequency and applying the method of averaging, if the displacement and force are approximated by harmonic functions of $\hat{t}$, $x(\hat{t})=a_x(\hat{t})\mathrm{sin}(\hat{\omega}\hat{t}+\phi_x(\hat{t}))$ and $g(\hat{t})=a_g(\hat{t})\mathrm{sin}(\hat{\omega}\hat{t}+\phi_g(\hat{t}))$, with $\hat{\omega}=\omega/\omega_0$, the slow dynamics of a Duffing non-linear damped resonator can be modeled by
\begin{equation}
    \dot{a}_x = -\zeta a_x - \beta\frac{a_x^3}{4} - \frac{1}{2} a_g \sin\phi \equiv \eta_a, \\[6pt]
\end{equation}
\begin{equation}
    \dot{\phi}_x = \frac{3}{8}\gamma a_x^{2} - (\hat{\omega}-1) - \frac{1}{2}\frac{a_g}{a_x}\cos\phi \equiv \eta_\phi, \\[6pt]
\end{equation}
\begin{equation}
    \phi = \phi_x - \phi_g.
\end{equation}
This non-linear system of differentiable equations correlates amplitude and phase dynamics. The system can be linearized to model the small perturbations around the operation point (set displacement amplitude and phase). The output of interest of the system are the small perturbations in displacement amplitude and phase, and are modeled through state space representation with displacement and force perturbations as follow
\begin{equation}
\begin{bmatrix}
\dot{\Delta a_x} \\
\dot{\Delta \phi_x}
\end{bmatrix}
=
\begin{bmatrix}
D_{11} & D_{12} \\
D_{21} & D_{22}
\end{bmatrix}
\begin{bmatrix}
\Delta a_x \\
\Delta \phi_x
\end{bmatrix}
+
\begin{bmatrix}
E_{11} & E_{12} \\
E_{21} & E_{22}
\end{bmatrix}
\begin{bmatrix}
\Delta a_g \\
\Delta \phi_g
\end{bmatrix}
\end{equation}
with the elements of the matrices given by
\begin{align}
D_{11}
    &= \left. \frac{\partial \eta_a}{\partial a_x} \right|_{a_{x0},\phi_0,a_{g0},\psi_0}
    = -\zeta - \frac{3 a_{x0}^2}{4}\,\beta, \\[6pt]
D_{12}
    &= \left. \frac{\partial \eta_a}{\partial \phi_x} \right|_{a_{x0},\phi_0,a_{g0},\psi_0}
    = \frac{a_{x0}}{\tan\phi_0}\left(\zeta + \frac{a_{x0}^2}{4}\beta\right), \\[6pt]
D_{21}
    &= \left. \frac{\partial \eta_\phi}{\partial a_x} \right|_{a_{x0},\phi_0,a_{g0},\psi_0}
    = \frac{3}{4}\gamma a_{x0}
      - \frac{1\zeta + \frac{a_{x0}^2}{4}\beta}{a_{x0}\tan\phi_0}, \\[6pt]
D_{22}
    &= \left. \frac{\partial \eta_\phi}{\partial \phi_x} \right|_{a_{x0},\phi_0,a_{g0},\psi_0}
    = -\left(\zeta + \frac{a_{x0}^2}{4}\beta\right), \\[6pt]
E_{11}
    &= \left. \frac{\partial \eta_a}{\partial a_g} \right|_{a_{x0},\phi_0,a_{g0},\psi_0}
    = -\frac{\sin\phi_0}{2}, \\[6pt]
E_{12}
    &= \left. \frac{\partial \eta_a}{\partial \phi_g} \right|_{a_{x0},\phi_0,a_{g0},\psi_0}
    = -\frac{a_{x0}}{\tan\phi_0}\left(\zeta + \frac{a_{x0}^2}{4}\beta\right), \\[6pt]
E_{21}
    &= \left. \frac{\partial \eta_\phi}{\partial a_g} \right|_{a_{x0},\phi_0,a_{g0},\psi_0}
    = -\frac{\cos\phi_0}{2a_{x0}}, \\[6pt]
E_{22}
    &= \left. \frac{\partial \eta_\phi}{\partial \phi_g} \right|_{a_{x0},\phi_0,a_{g0},\psi_0}
    = \zeta + \frac{a_{x0}^2}{4}\beta.
\end{align}

Taking the derivation to Laplace domain, the system is given by
\begin{equation}
\begin{bmatrix}
A_x(s) \\
\Phi_x(s)
\end{bmatrix}
=
\begin{bmatrix}
H_{11}(s) & H_{12}(s) \\
H_{21}(s) & H_{22}(s)
\end{bmatrix}
\begin{bmatrix}
A_g(s) \\
\Phi_g(s)
\end{bmatrix},
\end{equation}
where $A_x(s)$, $\Phi_x(s)$, $A_g(s)$, and $\Phi_g(s)$ are the Laplace transforms of 
$\Delta a_x$, $\Delta \phi_x$, $\Delta a_g$, and $\Delta \phi_g$, respectively.
The translation of the state-space representation to Laplace transfer functions is simply achieved by setting $H(s)=(s\mathrm{I_2}-D)^{-1}E$, where $I_2$ is the identity matrix. For simplification, the phase setpoint can be redefined as a constant $\phi_0=\pi/2$, as the objective is to drive the NMR at resonance. This yields the components of the transfer function matrix: 
\begin{align}
H_{11} &= -\,\frac{2}{\,3a^{2}\beta + 4\zeta + 4s\,}, \\[10pt]
H_{12} &= 0, \\[10pt]
H_{21} &= -\,\frac{6\gamma a}
{(\,3a^{2}\beta + 4\zeta + 4s\,)\;(\,a^{2}\beta + 4\zeta + 4s\,)}, \\[10pt]
H_{22} &= 
\frac{4\left(\zeta + \tfrac{1}{4}a^{2}\beta\right)}
{\,a^{2}\beta + 4\zeta + 4s\,}.
\end{align}
As mentioned in the main text, the vibration of the resonator is tracked and controlled through a PLL transduction scheme. Control loop passes the perturbations along with the feedback signal back to the resonator, which needs to be accounted for by a matrix $B(s)$ transferring perturbations of the output back to the input. Luckily, as shown by \cite{zhang_enhanced_2025} the intersections of frequency noise are independent of PLL parameters, therefore, for the problem at stake the parameters of the control loop can be selected for convenience. Noteworthy, when the PLL bandwidth is set to infinite, the phase perturbations are fed back with a gain of 1 \cite{demir_understanding_2021}. On the other hand, the feedback controller only detects and effects the phase signal, which means that any amplitude perturbations in or out of the feedback are ignored. Then, the feedback matrix is given by
\begin{equation}
    B(s)=
    \begin{bmatrix}
    0 & 0 \\
    0 & 1
\end{bmatrix}
\end{equation}
Incorporating the feedback, the close-loop transfer function matrix $J(s)$ describing the dynamics of phase and amplitude perturbation is then given by 
\begin{equation}
    J(s)=[I-H(s)B(s)]^{-1}H(s)
\end{equation}
The components of the matrix J(s) are then given by
\begin{align}
J_{11} &= 
-\frac{2}{3 a^{2}\beta + 4\zeta + 4 s}
, \\[10pt]
J_{12} &= 0, \\[10pt]
J_{21} &=
-\frac{3\,\gamma\,a}{2\,s\left(3 a^{2}\beta + 4\zeta + 4 s\right)}
, \\[10pt]
J_{22} &= 
\frac{\zeta + \tfrac{1}{4} a^{2}\beta}{s}
.
\end{align}
The power spectral density (PSD) of the control loop output amplitude and phase perturbations, $S_{a_x}(\omega)$ and $S_{\phi_x}(\omega)$, are then correlated with the PSD of the input amplitude and phase perturbations, $S_\mathrm{a|thm}$ and $S_\mathrm{\phi|thm}$ through 
\begin{equation}\label{Transfer Function}
\begin{bmatrix}
S_{a_x}(\omega) \\
S_{\phi_x}(\omega)
\end{bmatrix}
=
\begin{bmatrix}
|J_{11}(i\omega)|^2 & |J_{12}(i\omega)|^2 \\
|J_{21}(i\omega)|^2 & |J_{22}(i\omega)|^2 
\end{bmatrix}
\begin{bmatrix}
S_{a|\mathrm{thm}} \\
S_{\phi|\mathrm{thm}}
\end{bmatrix}.
\end{equation}
with $S_\mathrm{a|thm}$ and $S_\mathrm{\phi|thm}$ given by \cite{rubiola_phase_2008}
\begin{equation}
S_{\phi|\mathrm{thm}}
    = \frac{8 k_B T Q }{m \omega_0^3 a_x^2}
\end{equation}
and
\begin{equation}
S_\mathrm{a|\mathrm{thm}}
    = \frac{8 k_B T}{Q m \omega_0^3}.
\tag{18}
\end{equation}

With these input PSDs, the output amplitude and phase perturbation spectra,
$S_{a_x}$ and $S_{\phi_x}$, can be computed using \autoref{Transfer Function}.

Ultimately, the fractional frequency noise is the variable of interest, which is correlated to phase noise simply through $S_\mathrm{y}=S_{\phi_x}\left(\frac{\omega}{\omega_0}\right)^2$, yielding the result presented in the main text:
\begin{equation} 
\begin{aligned} 
S_\mathrm{y,\mathrm{thm}} ={}&\;
\left(\frac{\omega}{\omega_{0}}\right)^{2}
\Biggl(
\frac{
18\,\gamma^{2}a^{2}k_{B}T_{d}\left(2\zeta + 2a^{2}\beta\right)
}{
\omega^{2}\omega_{0}
\left(
\left(3a^{2}\beta + 4\zeta\right)^{2}
+ \dfrac{16\omega^{2}}{\omega_{0}^{2}}
\right)
m_{\mathrm{eff}}
}
\\[6pt]
&\quad
+
\frac{
\left(a^{2}\beta + 4\zeta\right)^{2}k_{B}T_{d}
}{
2\,\omega_{0}\,\omega^{2}
\left(2\zeta + 2a^{2}\beta\right)
m_{\mathrm{eff}}\,a^{2}
}
\Biggr) .
\end{aligned}
\end{equation}
The full form of thermomechanical fluctuation spectrum can also be decomposed into three approximate asymptotes for three distinct frequency ranges. The boundaries between the three ranges are given by 
\begin{equation}
    \omega_{\mathrm{c_{low}}}=
\frac{
    3\,a^{2}\gamma\,\omega_{0}
    \left(\zeta + a^{2}\beta\right)
    \left(4\zeta + 3a^{2}\beta\right)
}{
    \left(144\,a^{4}\gamma^{2}\left(\zeta + a^{2}\beta\right)^{2}
    + \left(4\zeta + a^{2}\beta\right)^{2}
      \left(4\zeta + 3a^{2}\beta\right)^{2}\right)^{1/2}
}
\end{equation}
and
\begin{equation}
    \omega_{\mathrm{c_{high}}}=
\frac{3\,\gamma\,a^{2}\,\omega_{0}\,\left(\zeta + a^{2}\beta\right)}
     {4\zeta + a^{2}\beta}
     \label{wchigh}
\end{equation}

The function is then defined as
\begin{equation}
\begin{aligned}
S_\mathrm{y}(\omega \ll \omega_{\mathrm{c_{low}}})
={}&
\frac{k_{B}T}{\omega_{0}^{3} m_{\mathrm{eff}}}
\Biggl(
\frac{36\,\gamma^{2} a^{2}\left(\zeta + a^{2}\beta\right)}
{\left(4\zeta + 3a^{2}\beta\right)^{2}}
\\
&\quad+
\frac{\left(4\zeta + a^{2}\beta\right)^{2}}
{4 a^{2}\left(\zeta + a^{2}\beta\right)}
\Biggr)
\end{aligned}
\end{equation}
\begin{equation}
S_\mathrm{y}(\omega_{\mathrm{c_{low}}} \ll \omega \ll \omega_{\mathrm{c_{high}}}) =\frac{
9\,\gamma^{2} a^{2} k_{B} T \left(\zeta + a^{2}\beta\right)
}{
4\,\omega_{0}\, m_{\mathrm{eff}}\, \omega^{2}
}
\end{equation}
\begin{equation}
S_\mathrm{y}(\omega \gg \omega_{\mathrm{c_{high}}}) =\frac{
\left(4\zeta + a^{2}\beta\right)^{2} k_{B} T
}{
4 a^{2} \omega_{0}^{3} \left(\zeta + a^{2}\beta\right) m_{\mathrm{eff}}
}
\end{equation}
\section{FEM Validation simulations of in-plane displacement}\label{FEM}
\autoref{Qnl} and \autoref{beta} in the main text are valid when dissipation from in-plane displacement is negligible relative to dissipation resulting from out-of-plane displacement. The finite element simulation (FEM) in this section demonstrate that this condition is satisfied for the square resonators considered in Sec. V. 

\begin{figure}
    \centering
    \includegraphics[width=1\linewidth]{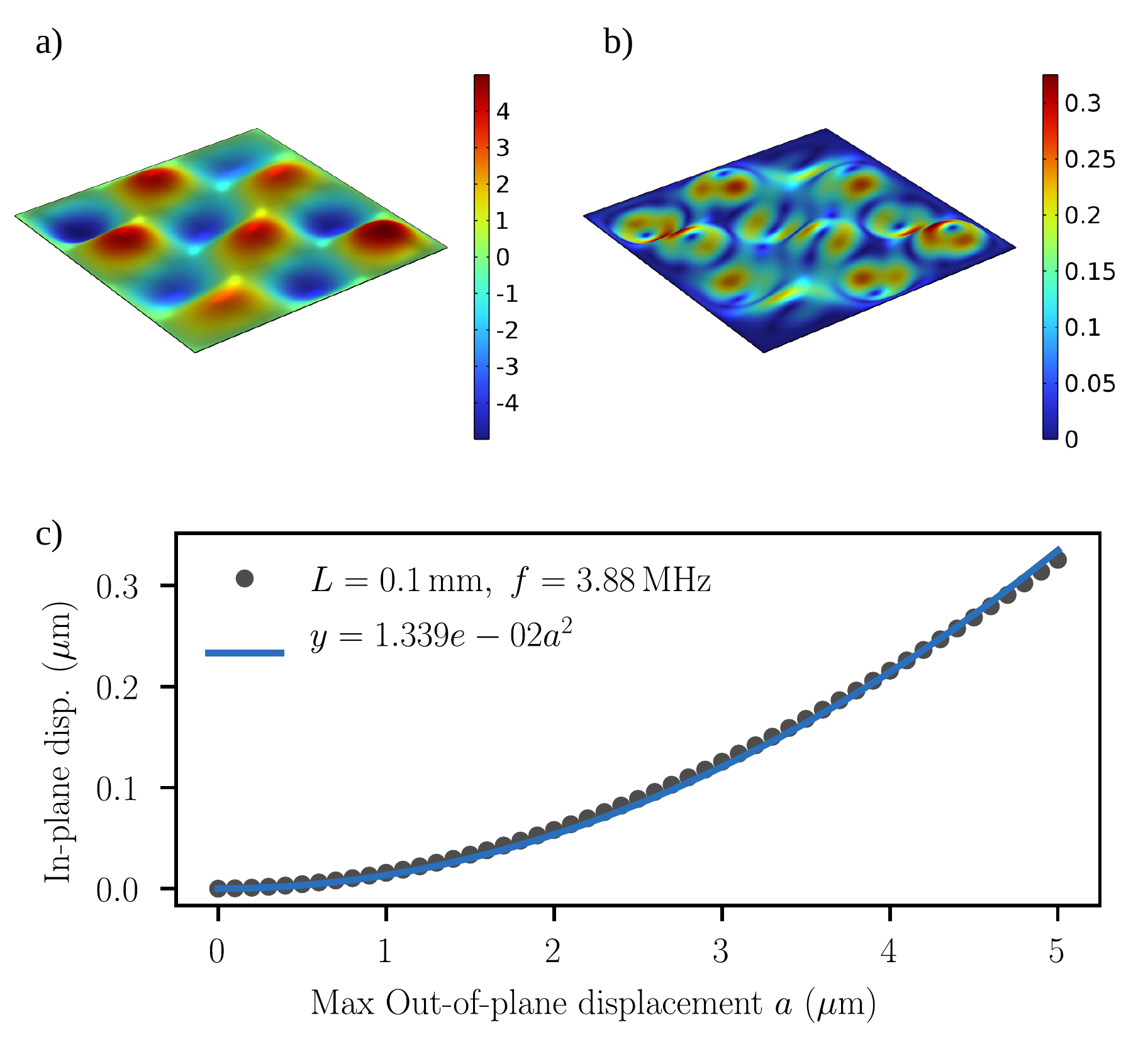}
    \caption{(a) Out of plane displacement in the worst case studied scenario and (b), corresponding in-place displacement. Both color scales are in units of $\mu\mathrm{m}$ (c) Quadratic relation between in-plane and out-of-plane displacement.}
    \label{fig:S1}
\end{figure}

We first extract the quadratic relation between the in-plane displacement and the out-of-plane displacement for square membrane resonators of various representative sizes (0.1, 1, and 3 mm), and for both the fundamental mode and a high order mode (mode order (4,3)). We do so by applying a constant out-of-plane displacement, following a given mode shape, and by tracking the resulting in-plane displacement. In a worst-case scenario (0.1 mm membrane, and (4,3) mode order, see \autoref{fig:S1}) we obtain the following relation between the maxiumum in-plane ($\Delta x$) and out-of-plane displacements ($a$): 
\begin{equation} \label{S1}
    \Delta x=1.339 \times 10^{-2} a^2.
\end{equation} 
Conversely, the least important in-plane displacement ($\Delta x=1.339 \times 10^{-4}a^2$) is obtained for the fundamental mode of a 3 mm membrane.

Considering a simple spring relation for in plane displacement, we expect the in-plane dissipated energy to be on the order of $\Delta W_{\parallel}\approx \frac{k_{\parallel}\Delta x^2}{2Q_{\mathrm{int}}}$ , with $k_{\parallel}=Eh$. Considering the eigenmode energy $W$, the Q-factor due to in-plane displacement ($Q_{\parallel}$) is therefore on the order of: 
\begin{equation}
    Q_{\parallel} = \frac{W}{\Delta W_{\parallel}} \approx \frac{\pi^2 \sigma q_{\mathrm{int}}\, h (n^2 + j^2)\, a^2}{4 E \Delta x^2}.
\end{equation}

In contrast, the Q-factor due to non-linear damping ($Q_{\mathrm{nl}}$), defined by \autoref{Qnl}) is many orders of magnitude lower. Indeed, even in our worst case scenario, the ratio between $Q_{\parallel}$ and $Q_{\mathrm{nl}}$ remains extremely large, independently of the drive amplitude $a$: 
\begin{equation} \label{S3}
    \frac{Q_{\parallel}}{Q_{\mathrm{nl}}}=\frac{1.05\times 10^4a^2}{3.23\times10^{-9}a^2}=3.2\times 10^{12}.
\end{equation}

We therefore conclude that the contribution of in-plane displacement to dissipation is negligible for square membrane resonators as long as the eigenmode orders are not extremely large. Given the extremely large magnitude of the ratio in \autoref{S3}, we expect the same assumption to hold for any membrane-like resonators (such as trampolines), except maybe for devices specifically designed to have a near-null duffing coefficient ($\gamma \approx 0$), such as \cite{Li2024}. 

\begin{figure}
    \centering
    \includegraphics[width=1\linewidth]{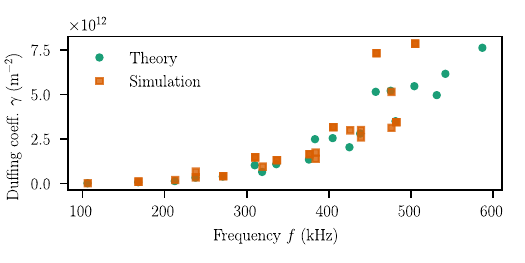}
    \caption{Comparison between the Duffing coefficients ($\gamma$) obtained analytically and using FEM simulations.}
    \label{fig:Duffing_plot}
\end{figure}

The data obtained from FEM simulations presented above was obtained using COMSOL Multiphysics. A mesh convergence analysis was done to validate the accuracy of the results used to obtain \autoref{S1}. For this analysis, we compared values of the Duffing coefficient ($\gamma$) obtained by our FEM simulations, to the analytical solution of \autoref{Duffing coefficient}. Numerically, $\gamma$ is obtained by computing the following surface integral–adapted from \cite{catalini_modeling_2021} over each mode shape: 
\begin{equation}
    \gamma = \frac{-hE}{2 m_{\mathrm{eff}} (1 - \nu^2)\omega_0^2} \int  
    \phi \left( \partial_{xy}\phi \, \partial_x \phi \, \partial_y \phi 
    + \frac{\nu}{1 - \nu} \, \partial_{xx}\phi \, \partial_{yy}\phi \right) dA
\end{equation}
where $\phi$ is the mode shape. An example results of this comparison is plotted in \autoref{fig:Duffing_plot} which shows a minimal difference between the Duffing coefficients obtained using our FEM simulations and the analytical model. 

\bibliographystyle{apsrev4-2}
\bibliography{apssamp}

@PREAMBLE{
 "\providecommand{\noopsort}[1]{}" 
 # "\providecommand{\singleletter}[1]{#1}%" 
}

@article{mather_bolometer_1982,
  title   = {Bolometer noise: nonequilibrium theory},
  author  = {Mather, John C.},
  journal = {Applied Optics},
  volume  = {21},
  number  = {6},
  pages   = {1125--1129},
  year    = {1982},
  doi     = {10.1364/AO.21.001125}
}

@phdthesis{misiak_developpements_2021,
  title  = {Développements de nouveaux détecteurs cryogéniques bas seuils pour la recherche de matière noire légère et la physique des neutrinos de basse énergie},
  author = {Misiak, Dimitri},
  school = {Université de Lyon},
  year   = {2021}
}

@article{catalini_modeling_2021,
  title   = {Modeling and Observation of Nonlinear Damping in Dissipation-Diluted Nanomechanical Resonators},
  author  = {Catalini, Letizia and Rossi, Massimiliano and Langman, Eric C. and Schliesser, Albert},
  journal = {Physical Review Letters},
  volume  = {126},
  number  = {17},
  pages   = {174101},
  year    = {2021},
  doi     = {10.1103/PhysRevLett.126.174101}
}

@article{demir_understanding_2021,
  title   = {Understanding fundamental trade-offs in nanomechanical resonant sensors},
  author  = {Demir, Alper},
  journal = {Journal of Applied Physics},
  volume  = {129},
  number  = {4},
  pages   = {044503},
  year    = {2021},
  doi     = {10.1063/5.0035254}
}

@article{manzaneque_resolution_2023,
  title   = {Resolution Limits of Resonant Sensors},
  author  = {Manzaneque, Tomás and Ghatkesar, Murali K. and Alijani, Farbod and Xu, Minxing and Norte, Richard A. and Steeneken, Peter G.},
  journal = {Physical Review Applied},
  volume  = {19},
  number  = {5},
  pages   = {054074},
  year    = {2023},
  doi     = {10.1103/PhysRevApplied.19.054074}
}

@article{zhang_enhanced_2025,
  title   = {Enhanced Bandwidth in Radiation Sensors Operating at the Fundamental Temperature Fluctuation Noise Limit},
  author  = {Zhang, Chang and Louis-Seize, Zachary and Saleh, Yahya and Brazeau, Maxime and Hodges, Timothy and Turgeon-Roy, Mathis and St-Gelais, Raphael},
  journal = {Nano Letters},
  volume  = {25},
  number  = {40},
  pages   = {14660--14667},
  year    = {2025},
  doi     = {10.1021/acs.nanolett.5c03415}
}

@book{Schmid2023Fundamentals,
  title     = {Fundamentals of Nanomechanical Resonators},
  author    = {Schmid, Silvan and Villanueva, Luis Guillermo and Roukes, Michael Lee},
  publisher = {Springer},
  address   = {Cham},
  year      = {2023},
  edition   = {2},
  doi       = {10.1007/978-3-031-29628-4}
}

@article{kanellopulos_comparative_2025,
  title   = {Comparative analysis of nanomechanical resonators: sensitivity, response time, and practical considerations in photothermal sensing},
  author  = {Kanellopulos, Kostas and Ladinig, Friedrich and Emminger, Stefan and Martini, Paolo and West, Robert G. and Schmid, Silvan},
  journal = {Microsystems \& Nanoengineering},
  volume  = {11},
  number  = {1},
  pages   = {28},
  year    = {2025},
  doi     = {10.1038/s41378-025-00879-6}
}

@article{zhang_high_2024,
  title   = {High detectivity terahertz radiation sensing using frequency-noise-optimized nanomechanical resonators},
  author  = {Zhang, Chang and Yalavarthi, Eeswar K. and Giroux, Mathieu and Cui, Wei and Stephan, Michel and Maleki, Ali and Weck, Arnaud and M{\'e}nard, Jean-Michel and St-Gelais, Raphael},
  journal = {APL Photonics},
  volume  = {9},
  number  = {12},
  pages   = {126105},
  year    = {2024},
  doi     = {10.1063/5.0238977}
}

@incollection{lifshitz_nonlinear_2008,
  title     = {Nonlinear dynamics of nanomechanical and micromechanical resonators},
  author    = {Lifshitz, Ron and Cross, Michael C.},
  booktitle = {Reviews of Nonlinear Dynamics and Complexity},
  volume    = {1},
  pages     = {1--52},
  year      = {2008},
  publisher = {Wiley-VCH},
  address   = {Weinheim}
}

@article{zhang_nanomechanical_2013,
  title   = {Nanomechanical torsional resonators for frequency-shift infrared thermal sensing},
  author  = {Zhang, X. and Myers, E. and Sader, J. and Roukes, M.},
  journal = {Nano Letters},
  year    = {2013},
  volume  = {13},
  pages   = {1528--1534}
}

@article{laurent_electromechanical_2018,
  title   = {12-$\mu$m-pitch electromechanical resonator for thermal sensing},
  author  = {Laurent, L. and Yon, J.-J. and Moulet, J.-S. and Roukes, M. and Duraffourg, L.},
  journal = {Physical Review Applied},
  year    = {2018},
  volume  = {9},
  pages   = {024016}
}

@article{blaikie_graphene_2019,
  title   = {A fast and sensitive room-temperature graphene nanomechanical bolometer},
  author  = {Blaikie, A. and Miller, D. and Alem{\'a}n, B. J.},
  journal = {Nature Communications},
  year    = {2019},
  volume  = {10},
  pages   = {4726}
}

@article{hui_plasmonic_2016,
  title   = {Plasmonic piezoelectric nanomechanical resonator for spectrally selective infrared sensing},
  author  = {Hui, Y. and Gomez-Diaz, J. S. and Qian, Z. and Alu, A. and Rinaldi, M.},
  journal = {Nature Communications},
  year    = {2016},
  volume  = {7},
  pages   = {11249}
}

@article{piller_thermal_2023,
  title   = {Thermal IR detection with nanoelectromechanical silicon nitride trampoline resonators},
  author  = {Piller, M. and Hiesberger, J. and Wistrela, E. and Martini, P. and Luhmann, N. and Schmid, S.},
  journal = {IEEE Sensors Journal},
  year    = {2023},
  volume  = {23},
  pages   = {1066--1071}
}

@article{zhang_room_2016,
  title   = {Room temperature, very sensitive thermometer using a doubly clamped microelectromechanical beam resonator for bolometer applications},
  author  = {Zhang, Y. and Watanabe, Y. and Hosono, S. and Nagai, N. and Hirakawa, K.},
  journal = {Applied Physics Letters},
  year    = {2016},
  volume  = {108},
  pages   = {163503}
}

@article{vicarelli_micromechanical_2022,
  title   = {Micromechanical bolometers for subterahertz detection at room temperature},
  author  = {Vicarelli, L. and Tredicucci, A. and Pitanti, A.},
  journal = {ACS Photonics},
  year    = {2022},
  volume  = {9},
  pages   = {360--367}
}

@article{wang_vanadium_2012,
  title   = {Vanadium oxide microbolometer with gold black absorbing layer},
  author  = {Wang, B. and Lai, J. and Zhao, E. and Hu, H. and Liu, Q. and Chen, S.},
  journal = {Optical Engineering},
  year    = {2012},
  volume  = {51},
  pages   = {074003}
}

@article{renoux_subwavelength_2011,
  title   = {Sub-wavelength bolometers: Uncooled platinum wires as infrared sensors},
  author  = {Renoux, P. and J{\'o}nsson, S. {\AE}. and Klein, L. J. and Hamann, H. F. and Ingvarsson, S.},
  journal = {Optics Express},
  year    = {2011},
  volume  = {19},
  pages   = {8721--8727}
}

@article{hossain_pyroelectric_1991,
  title   = {Pyroelectric detectors and their applications},
  author  = {Hossain, A. and Rashid, M. H.},
  journal = {IEEE Transactions on Industry Applications},
  year    = {1991},
  volume  = {27},
  pages   = {824--829}
}

@article{liu_pyroelectric_1978,
  title   = {Pyroelectric detectors and materials},
  author  = {Liu, S. and Long, D.},
  journal = {Proceedings of the IEEE},
  year    = {1978},
  volume  = {66},
  pages   = {14--26}
}

@article{hsu_graphene_2015,
  title   = {Graphene-based thermopile for thermal imaging applications},
  author  = {Hsu, A. L. and Herring, P. K. and Gabor, N. M. and Ha, S. and Shin, Y. C. and Song, Y. and Chin, M. and Dubey, M. and Chandrakasan, A. P. and Kong, J. and others},
  journal = {Nano Letters},
  year    = {2015},
  volume  = {15},
  pages   = {7211--7216}
}

@article{stantoine_nanotube_2011,
  title   = {Single-walled carbon nanotube thermopile for broadband light detection},
  author  = {St-Antoine, B. C. and M{\'e}nard, D. and Martel, R.},
  journal = {Nano Letters},
  year    = {2011},
  volume  = {11},
  pages   = {609--613}
}

@book{kruse_elements_1962,
  title     = {Elements of Infrared Technology: Generation, Transmission, and Detection},
  author    = {Kruse, Paul W. and McGlauchlin, Laurence D. and McQuistan, Richmond B.},
  publisher = {Wiley},
  address   = {New York},
  year      = {1962},
  isbn      = {9780471508861}
}

@article{woltersdorff_uber_1934,
  title   = {Über die optischen Konstanten dünner Metallschichten im langwelligen Ultrarot},
  author  = {Woltersdorff, W.},
  journal = {Zeitschrift für Physik},
  year    = {1934},
  volume  = {91},
  pages   = {230--252}
}

@article{hadley_reflection_1947,
  title   = {Reflection and transmission interference filters},
  author  = {Hadley, L. N. and Dennison, D. M.},
  journal = {Journal of the Optical Society of America},
  year    = {1947},
  volume  = {37},
  pages   = {451--465}
}

@article{hilsum_infrared_1955,
  title   = {Infrared absorption of thin metal films at non-normal incidence},
  author  = {Hilsum, C.},
  journal = {Journal of the Optical Society of America},
  year    = {1955},
  volume  = {45},
  pages   = {135--136}
}

@article{luhmann_ultrathin_2020,
  title   = {Ultrathin 2 nm gold as impedance-matched absorber for infrared light},
  author  = {Luhmann, Niklas and H{\o}j, Dennis and Piller, Markus and K{\"a}hler, Hendrik and Chien, Miao-Hsuan and West, Robert G. and Andersen, Ulrik Lund and Schmid, Silvan},
  journal = {Nature Communications},
  year    = {2020},
  volume  = {11},
  pages   = {2161},
  doi     = {10.1038/s41467-020-15762-3}
}

@article{villanueva_evidence_2014,
  title   = {Evidence of Surface Loss as Ubiquitous Limiting Damping Mechanism in SiN Micro- and Nanomechanical Resonators},
  author  = {Villanueva, L. G. and Schmid, S.},
  journal = {Physical Review Letters},
  year    = {2014},
  volume  = {113},
  number  = {22},
  pages   = {227201},
  doi     = {10.1103/PhysRevLett.113.227201}
}

@article{zink2004specific,
  title={Specific heat and thermal conductivity of low-stress amorphous silicon nitride},
  author={Zink, B. L. and Hellman, F.},
  journal={Solid State Communications},
  volume={129},
  number={3},
  pages={199--204},
  year={2004},
  publisher={Elsevier}
}

@article{rugar_improved_1989,
  title   = {Improved fiber-optic interferometer for atomic force microscopy},
  author  = {Rugar, D. and Mamin, H. J. and Guethner, P.},
  journal = {Applied Physics Letters},
  year    = {1989},
  volume  = {55},
  number  = {25},
  pages   = {2588--2590},
  doi     = {10.1063/1.101987}
}

@article{shin_spiderweb_2022,
  title   = {Spiderweb Nanomechanical Resonators via Bayesian Optimization: Inspired by Nature and Guided by Machine Learning},
  author  = {Shin, Dongil and Cupertino, Andrea and de Jong, Matthijs H. J. and Steeneken, Peter G. and Bessa, Miguel A. and Norte, Richard A.},
  journal = {Advanced Materials},
  year    = {2022},
  volume  = {34},
  number  = {3},
  pages   = {2106248},
  doi     = {10.1002/adma.202106248}
}

@article{bereyhi_hierarchical_2022,
  title   = {Hierarchical tensile structures with ultralow mechanical dissipation},
  author  = {Bereyhi, M. J. and Beccari, A. and Groth, R. and Fedorov, S. A. and Arabmoheghi, A. and Kippenberg, T. J. and Engelsen, N. J.},
  journal = {Nature Communications},
  year    = {2022},
  volume  = {13},
  pages   = {3097},
  doi     = {10.1038/s41467-022-30586-z}
}

@article{lu_nonlinear_2020,
  title   = {Nonlinear vibration control effects of membrane structures with in-plane PVDF actuators: A parametric study},
  author  = {Lu, Y. and Shao, Q. and Amabili, M. and Yue, H. and Guo, H.},
  journal = {International Journal of Non-Linear Mechanics},
  year    = {2020},
  volume  = {122},
  pages   = {103466},
  doi     = {10.1016/j.ijnonlinmec.2020.103466}
}

@book{rubiola_phase_2008,
  title     = {Phase Noise and Frequency Stability in Oscillators},
  author    = {Rubiola, Enrico},
  publisher = {Cambridge University Press},
  year      = {2008}
}

@article{Underwood2015,
  title = {Measurement of the motional sidebands of a nanogram-scale oscillator in the quantum regime},
  author = {Underwood, M. and Mason, D. and Lee, D. and Xu, H. and Jiang, L. and Shkarin, A.~B. and B{\o}rkje, K. and Girvin, S.~M. and Harris, J.~G.~E.},
  journal = {Physical Review A},
  volume = {92},
  number = {6},
  pages = {061801(R)},
  year = {2015},
  doi = {10.1103/PhysRevA.92.061801},
  url = {https://link.aps.org/doi/10.1103/PhysRevA.92.061801}
}

@article{He2026,
  title = {Uncooled infrared imaging using a single trampoline‐based optomechanical thermal detector},
  author = {He, Zi‐Han and Peng, Zhao‐Xin and Shi, Yu‐Hao and Neergaard‐Nielsen, Jonas Schou and Gong, Zhi‐Cheng and Chen, Xue‐Ying and Wu, Lin‐Qian and Si, Qi‐Hang and Andersen, Ulrik Lund and Li, Yong and Fu, Hao},
  journal = {Photonics Research},
  volume = {14},
  number = {2},
  pages = {587--594},
  year = {2026},
  doi = {10.1364/PRJ.579086},
  url = {https://opg.optica.org/prj/fulltext.cfm?uri=prj-14-2-587}
}

@article{Li2024,
  author  = {Li, Z. and Xu, M. and Norte, R. and Aragon, A. and Arabmoheghi, A. and Steeneken, P. and Alijani, F.},
  title   = {Strain engineering of nonlinear nanoresonators from hardening to softening},
  journal = {Communications Physics},
  volume  = {7},
  pages   = {53},
  year    = {2024},
  doi     = {10.1038/s42005-024-01543-7},
  url     = {https://doi.org/10.1038/s42005-024-01543-7}
}
\end{document}